\documentclass[a4paper,fleqn,usenatbib,useAMS]{mnras}

\usepackage{graphicx}	
\usepackage{amsmath}	
\usepackage{amssymb}	
\usepackage{multicol}        
\usepackage{bm}		
\usepackage{pdflscape}	
\usepackage{multirow}
\usepackage{xcolor}
\usepackage{hyperref}
\usepackage{booktabs}
\usepackage{subfigure}
\usepackage{makecell}
\usepackage{tikz}
\usepackage[normalem]{ulem}
\usepackage{fontawesome}

\newcommand{\kms}{\,km\,s$^{-1}$} 
\def\specialname[#1]{\textbf{\textsc{#1}}}

\definecolor{lime}{HTML}{A6CE39}
\DeclareRobustCommand{\orcidicon}{%
	\begin{tikzpicture}
	\draw[lime, fill=lime] (0,0)
	circle [radius=0.16]
	node[white] {{\fontfamily{qag}\selectfont \tiny ID}};
	\draw[white, fill=white] (-0.0625,0.095)
	circle [radius=0.007];
	\end{tikzpicture}
	\hspace{-2mm}
}
\foreach \x in {A, ..., Z}{%
	\expandafter\xdef\csname orcid\x\endcsname{\noexpand\href{https://orcid.org/\csname orcidauthor\x\endcsname}{\noexpand\orcidicon}}
}

\usepackage[T1]{fontenc}
\usepackage{ae,aecompl}

\usepackage{newtxtext,newtxmath}

\title[Estimate halo concentration]{
An efficient and robust method to estimate halo concentration based on the method of moments
}
\author[Wang et al.]{Kai Wang\orcidK{},$^{1}$\thanks{Contact e-mail:
    wkcosmology@gmail.com}
	H.J. Mo,$^{2}$
    Yangyao Chen\orcidC{}$^{3, 4}$
    and Joop Schaye\orcidJ{}$^{5}$
    \\
    $^1$Kavli Institute for Astronomy and Astrophysics, Peking University, Beijing 100871, China\\
	$^2$Department of Astronomy, University of Massachusetts Amherst, MA 01003, USA\\
    $^3$School of Astronomy and Space Science, University of Science and Technology of China, Hefei, Anhui 230026, China\\
    $^4$Key Laboratory for Research in Galaxies and Cosmology, Department of Astronomy, University of Science and Technology of China, Hefei, Anhui 230026, China \\
    $^5$Leiden Observatory, Leiden University, PO Box 9513, 2300 RA Leiden, the Netherlands
}

\date{Last updated 2020 May 22; in original form 2018 September 5}

\pubyear{2020}

\begin{document}
	\label{firstpage}
	\pagerange{\pageref{firstpage}--\pageref{lastpage}}
	\maketitle

\begin{abstract}
    We propose an efficient and robust method to estimate the halo
    concentration based on the first moment of the density distribution, which
    is $R_1\equiv \int_0^{r_{\rm vir}}4\pi r^3\rho(r)dr/M_{\rm vir}/r_{\rm
    vir}$. We find that $R_1$ has a monotonic relation with the concentration
    parameter of the NFW profile, and that a cubic polynomial function can fit
    the relation with an error $\lesssim 3\%$. Tests on ideal NFW halos show
    that the conventional NFW profile fitting method and the $V_{\rm
    max}/V_{\rm vir}$ method produce biased halo concentration estimation by
    $\approx 10\%$ and $\approx 30\%$, respectively, for halos with 100
    particles. In contrast, the systematic error for our $R_1$ method is
    smaller than 0.5\% even for halos containing only 100 particles.
    Convergence tests on realistic halos in $N$-body simulations show that the
    NFW profile fitting method underestimates the concentration parameter for
    halos with $\lesssim 300$ particles by $\gtrsim 20\%$, while the error for
    the $R_1$ method is $\lesssim 8\%$. We also show other applications of
    $R_1$, including estimating $V_{\rm max}$ and the Einasto concentration
    $c_{\rm e}\equiv r_{\rm vir}/r_{-2}$. The calculation of $R_1$ is efficient
    and robust, and we recommend including it as one of the halo properties in
    halo catalogs of cosmological simulations.
\end{abstract}


\section{Introduction}%
\label{sec:introduction}

Dark matter halos, as the building blocks of the cosmic structures in our
Universe, are virialized objects formed by gravitational instability. The
assembly of halos proceeds hierarchically, where small halos are formed early
and merge with each other to form larger ones. The assembly history of dark
matter halos correlates strongly to the halo structure, and many
semi-analytical models have been proposed to explain this correlation and to
predict halo structure from its assembly history
\citep[e.g.][]{navarroUniversalDensityProfile1997,
    wechslerConcentrationsDarkHalos2002, zhaoGrowthStructureDark2003,
    luOriginColdDark2006, zhaoACCURATEUNIVERSALMODELS2009,
correaAccretionHistoryDark2015, diemerAccuratePhysicalModel2019}. In addition,
galaxies are born in the centers of dark matter halos and evolve following the
halo assembly process \citep[see][for a
review]{moGalaxyFormationEvolution2010}, so that dark matter halos and galaxies
are tightly related to each other. This motivates many attempts to model the
galaxy-halo connection to understand galaxy formation and evolution
\citep[see][for reviews]{baughPrimerHierarchicalGalaxy2006,
moGalaxyFormationEvolution2010, wechslerConnectionGalaxiesTheir2018}. Clearly,
it is of paramount importance to accurately characterize the structure of dark
matter halos.

Numerical $N$-body simulations with collisionless cold dark matter particles
provide an essential tool to study the structure of dark matter halos. To begin
with, dark matter halos are identified with a halo-finding algorithm, such as
the Friends-of-Friends (FoF) algorithm
\citep[e.g.][]{huchraGroupsGalaxiesNearby1982,
davisEvolutionLargescaleStructure1985}. Based on the spherical collapse model,
a dark matter halo is defined as the collection of particles within a radius
within which the mean density reaches some chosen value. This radius is usually
referred to as the virial radius, $r_{\rm vir}$, and the total mass enclosed is
the halo mass, $M_{\rm vir}$. The radial density profiles of dark matter halos
are found to be well described by the universal NFW function,
\begin{equation}
    \rho(r) = \frac{\rho_0}{r/r_{\mathrm s}(1 + r/r_{\mathrm s})^2}\,,
    \label{eq:nfw}
\end{equation}
specified by the two free parameters, $r_{\mathrm s}$ and $\rho_0$, or
equivalently, $M_{\rm vir}$ and the halo concentration, $c\equiv r_{\rm
vir}/r_{\mathrm s}$ \citep{navarroUniversalDensityProfile1997}.

However, the determination of the concentration parameter for simulated halos
is not straightforward. Many methods have been used to estimate the
concentration parameters of simulated halos from the spatial distribution of
dark matter particles \citep[e.g.][]{jingDensityProfileEquilibrium2000,
    klypinResolvingStructureCold2001, bullockProfilesDarkHaloes2001,
    wechslerConcentrationsDarkHalos2002, zhaoMassRedshiftDependence2003,
    zhaoGrowthStructureDark2003, duffyDarkMatterHalo2008a,
    zhaoACCURATEUNIVERSALMODELS2009, klypinDarkMatterHalos2011,
klypinMultiDarkSimulationsStory2016}. One approach is to sample the radial
density distribution of simulated halos in discrete bins and fit it with the
NFW profile \citep[e.g.][]{bhattacharyaDarkMatterHalo2013}. This method has
several shortcomings. Firstly, the estimated concentration is subject to the
choice of discrete bins. The use of large bin sizes tends to smooth the
gradient of the radial profile and to cause an underestimate of the halo
concentration, while the use of too small bins can introduce too much noise. In
general, it is difficult to find an optimal binning strategy, particularly when
a halo is only sparsely sampled. Secondly, the fitting method relies on the
prior choice of the halo profile, which is the NFW profile in this case.
Therefore, any deviations from the NFW profile will make the output
concentration biased \citep{einastoConstructionCompositeModel1965,
navarroInnerStructureLCDM2004, wangUniversalStructureDark2020}. Finally, the
fitting procedure is relatively time-consuming, making it difficult to do for
the large number of halos found in large cosmological simulations.

To overcome some of the issues in the NFW profile fitting, other estimators of
halo concentration have been proposed. One example is the method based on
$V_{\rm max}/V_{\rm vir}$ \citep{klypinResolvingStructureCold2001,
    klypinDarkMatterHalos2011, pradaHaloConcentrationsStandard2012,
klypinMultiDarkSimulationsStory2016}, where $V_{\rm
max}=\texttt{MAX}(\sqrt{GM(<r)/r})$ is the maximum of the circular velocity as
a function of $r$, and $V_{\rm vir}=\sqrt{GM(<r_{\rm vir})/r_{\rm vir}}$ is the
virial velocity. This quantity is closely related to the halo concentration for
an NFW halo. However, this relation is ill-defined for halos with concentration
below 2.16 \citep{klypinResolvingStructureCold2001}, since by definition
$V_{\rm max}$ cannot be smaller than $V_{\rm vir}$. The halo concentration can
also be inferred from $r_{\mathrm f}/r_{\rm vir}$, where $r_{\mathrm f}$ is the
radius within which the mass is $fM_{\rm vir}$, with $0 < f<1$
\citep[e.g.][]{langVoronoiTessellationNonparametric2015}. In addition, there
are also methods based on the integrated mass profile
\citep{poveda-ruizQuantifyingControllingBiases2016} and the Voronoi
Tessellation \citep{langVoronoiTessellationNonparametric2015}.

In this paper, we propose an efficient and robust method to estimate the
concentration parameter of NFW halos based on the first moment of the density
distribution. \S\,\ref{sec:data} introduces the data we use in this study.
\S\,\ref{sec:method} introduces three different methods to estimate the halo
concentration, and we test their performance in \S\,\ref{sec:performance}.
\S\,\ref{sec:application_to_elucid} shows the mass-concentration relation
obtained from the ELUCID simulation, to demonstrate the application of our
method to large cosmological simulations.
\S\,\ref{sec:other_applications_of_r_1_} discusses other applications of $R_1$,
including estimating $V_{\rm max}$ of halos and the Einasto concentration
parameter. Finally, \S\,\ref{sec:summary} summarizes our main results.
Throughout this paper, we use ``log'' to denote 10-based logarithm.

\section{Data}%
\label{sec:data}

The estimation of halo concentration is subject to the sampling effect, where
low-mass halos are poorly sampled in $N$-body simulations, and the force
softening effect, which can smooth the density profile in the inner region and
cause an underestimation of the halo concentration
\citep[e.g.][]{powerInnerStructureLCDM2003,
ludlowNumericalConvergenceSimulations2019, mansfieldHowBiasedAre2021}. To
separate the impact of these two effects, we use two different datasets to test
the performances of different methods: ideal halos generated from NFW profiles
with different halo parameters, and realistic halos selected from $N$-body
simulations of different resolutions and force softening lengths. We also apply
our method to a large $N$-body simulation to demonstrate its capability of
recovering halo concentrations in large cosmological simulations.

\subsection{Ideal NFW halos}%
\label{sub:idea_nfw_halo}

We generate ideal NFW halos using the \textsc{HaloFactory} package based on
Eddington's inversion method (see appendix~\ref{sec:halo_factory} for details).
Here we use halos with four different concentrations, $c=1$, $5$, $10$, and
$20$, a range sufficient to cover halos in cosmological $N$-body simulations.
For each concentration, we generate individual halos using different numbers of
particles, ranging from $\sim 100$ to $\gtrsim 10^4$, to test the robustness of
a given concentration estimator. For a given particle number, we use 10,000
random realizations to evaluate the statistical uncertainties.

\subsection{$N$-body simulations}%
\label{sub:n_body_simulation}

\begin{table*}
    \centering
	\caption{Summary of the $N$-body simulations used in this study.}
	\label{tab:simulation}
	\begin{tabular}{c|c|c|c|c}
		\hline
        Simulation & $L_{\rm box}$ $[h^{-1}{\rm Mpc}]$ & Particle number & Particle mass $[h^{-1}{\rm M_\odot}]$  & Gravitational softening length $[h^{-1}{\rm kpc}]$\\
		\hline
		\hline
        TNG50-1-Dark & $~~35$ & $2160^3$ & $3.7\times 10^5$  & 0.2\\
		\hline
        TNG100-1-Dark & $~~75$ & $1820^3$ & $6.0\times 10^6$  & 0.5\\
		\hline
        TNG100-3-Dark & $~~75$ & $~~455^3$ & $3.8\times 10^8$ & 2.0\\
		\hline
        ELUCID & 500 & $3072^3$ & $3.1\times 10^8$  & 3.5\\
		\hline
	\end{tabular}
\end{table*}

\subsubsection{IllustrisTNG-Dark}%
\label{ssub:illustristng}

The IllustrisTNG project consists of several dark-matter-only and
hydrodynamical simulations \citep{Pillepich_2018b, Nelson_2019}. Here we use
the dark-matter-only simulations with different resolutions and gravitational
softening lengths to test the impact on the concentration estimation. The
information of these simulations is summarized in Table~\ref{tab:simulation}
\citep{Nelson_2019}. It is noteworthy that TNG100-1-Dark and TNG100-3-Dark have
identical initial conditions, but different mass resolutions and gravitational
softening lengths. The IllustrisTNG project was based on a cosmology consistent
with the results in \citet{planckcollaborationPlanck2015Results2016}, where
$\Omega_\Lambda=0.6911$, $\Omega_{\mathrm m}=0.3089$, $\sigma_8=0.8159$,
$n_{\mathrm s}=0.9667$, and $h=0.6774$. Dark matter halos are
    identified using the Friends-of-Friends algorithm with a linking length
    that is 0.2 the mean inter-particle distance
\citep{davisEvolutionLargescaleStructure1985}, and their masses are
assigned as the total dark matter mass enclosed within the aperture where
the mean overdensity is 200 times the critical density. This mass is
denoted as $M_{\rm 200c}$, and the corresponding radius and concentration
are denoted as $r_{\rm 200c}$ and $c_{\rm 200c}$, respectively. The halo
center is specified as the location of the particle with the minimal
gravitational potential. Substructures are identified with the
\textsc{SUBFIND} algorithm \citep{springelPopulatingClusterGalaxies2001}.

\subsubsection{ELUCID}%
\label{ssub:elucid}

The ELUCID\footnote{\url{https://www.elucid-project.com/}} simulation
\citep{wangRECONSTRUCTINGINITIALDENSITY2013, wangELUCIDEXPLORINGLOCAL2014,
wangELUCIDEXPLORINGLOCAL2016, tweedELUCIDExploringLocal2017} is a constrained
simulation, run with a memory-optimized version of \textsc{GADGET-2}
\citep{springelSimulationsFormationEvolution2005} known as \textsc{L-GADGET},
to reconstruct the density field and formation history of our local Universe
based on the Sloan Digital Sky Survey DR7 \citep{yorkSloanDigitalSky2000}. It
is thus one particular realization of the structure formation model in
question. This simulation has $3072^3$ dark matter particles, each with a mass
of $3.09\times 10^8h^{-1}\rm M_\odot$, in a box with a side length of
$500h^{-1}\rm Mpc$. This simulation assumes a $\Lambda$CDM cosmology with
$\Omega_{\mathrm m} = 0.258$, $\Omega_\Lambda=0.742$, $\sigma_8=0.80$,
$n_{\mathrm s}=0.96$, and $h = 0.72$. The information of the ELUCID simulation
is summarized in Table~\ref{tab:simulation}
\citep{wangELUCIDEXPLORINGLOCAL2016}. ELUCID uses the same procedure as
IllustrisTNG to identify and define dark matter halos (see
\S\,\ref{ssub:illustristng}). The large volume and relatively high
    resolution of the ELUCID simulation allow us the investigate the
mass-concentration relation over a large halo mass range.

\section{Method}%
\label{sec:method}

Here we introduce three methods to estimate halo concentration: two commonly
used methods and our $R_1$ method. In addition, three other methods are
discussed in Appendix~\ref{sec:other_methods_to_estimate_halo_concentration}
together with their performance on ideal NFW halos.

\subsection{The $R_1$ method}%
\label{sub:the_r_1_method}

\begin{figure*}
    \centering
    \includegraphics[width=1\linewidth]{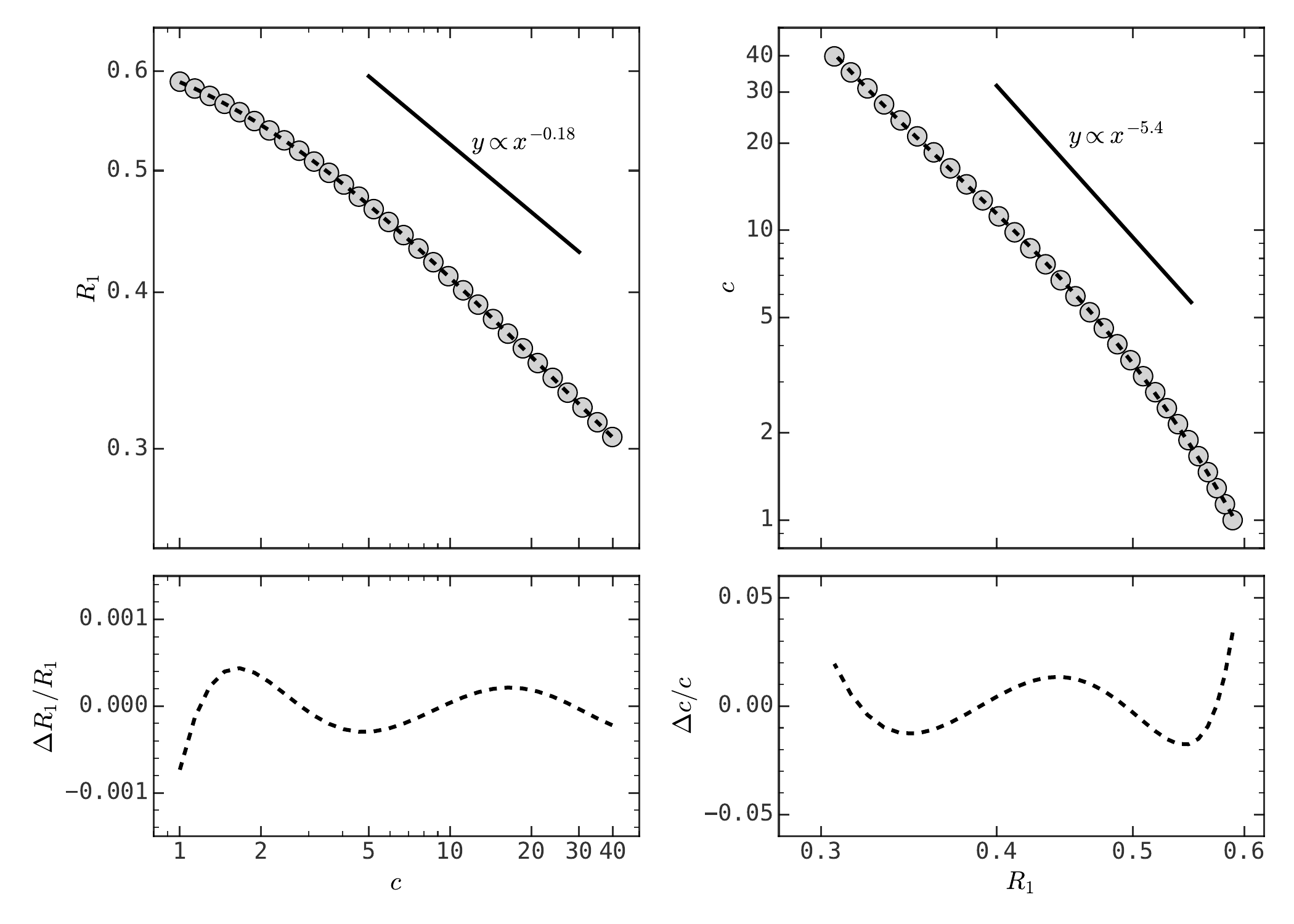}
    \caption{
        The relation between $c$ and $R_1$ for an NFW profile. Top panels:~The
        circles are the analytical results obtained through
        equation~(\ref{eq:r1_c}), and the dashed lines are the fitted
        third-order polynomial functions in equation~(\ref{eq:fitting}). Bottom
        panels: The fractional difference between the relation in
        equation~(\ref{eq:r1_c}) and the third-order polynomial fits.
    }%
    \label{fig:figures/conc_r1_relation}
\end{figure*}

\begin{figure*}
    \centering
    \includegraphics[width=1\linewidth]{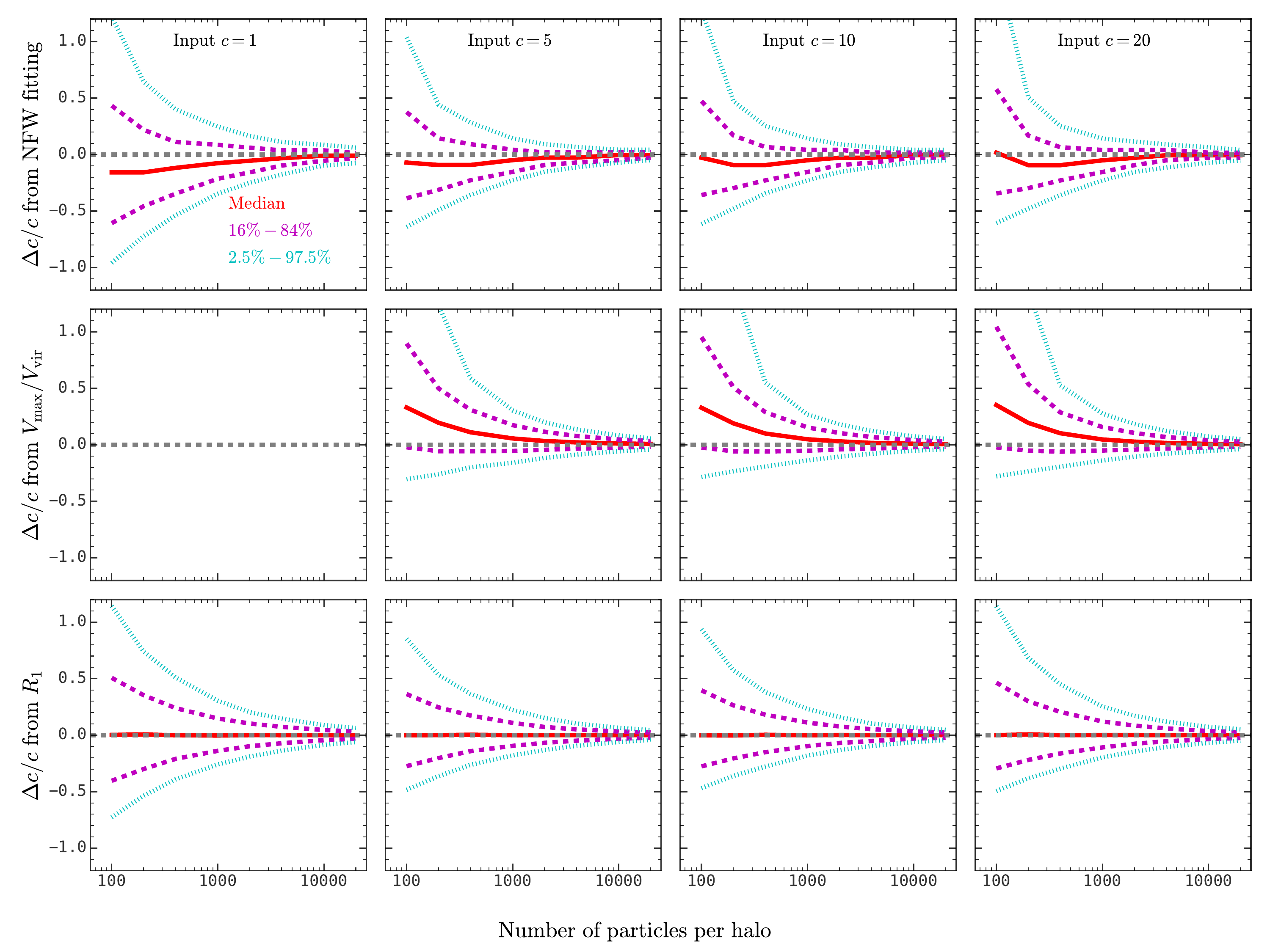}
    \caption{
        The fractional difference between the input and the estimated
        concentrations with the NFW fitting method (top panels), the $V_{\rm
        max}/V_{\rm vir}$ method (middle panels), and the $R_1$ method (bottom
        panels) for ideal NFW halos generated with \textsc{HaloFactory} as a
        function of the number of particles in the halo. The red solid, magenta
        dashed, and cyan dotted lines show the $50^{\rm th}$, $16^{\rm
        th}-84^{\rm th}$, and $2.5^{\rm th}-97.5^{\rm th}$ percentiles. This
        figure shows that the $R_1$ method gives an unbiased and less uncertain
        estimation of the input halo concentration compared with the other two
        methods. The middle left panel is empty since the $V_{\rm max}/V_{\rm
        vir}$ method cannot be applied to halos with $c<2.16$.
    }%
    \label{fig:figures/conc_ideal_test}
\end{figure*}

The total mass of a dark matter halo is expressed as
\begin{equation}
    M_{\rm vir} = \int_0^{r_{\rm vir}}4\pi r^2 \rho(r)dr\,.
\end{equation}
Here $\rho(r)$ is the radial density profile and $r_{\rm vir}$ is the halo
radius, which is usually defined as the radius within which the enclosed mean
density just exceeds some chosen value. The dimensionless first moment of the
density distribution, $R_1$, can be defined as
\begin{equation}
    R_1 = {1\over M_{\rm vir}r_{\rm vir}}\int_0^{r_{\rm vir}} 4\pi
    r^3\rho(r)dr\,, \label{eq:r1}
\end{equation}
which can be expressed analytically for an NFW profile as
\begin{equation}
    R_1 = \frac{c - 2 \ln(1 + c) + c/(1 +c)}{c\left[\ln(1 + c) - c/(1 +
    c)\right]}\,. \label{eq:r1_c}
\end{equation}
Despite the complicated functional form, the relation between $R_1$ and $c$
is actually quite simple, as shown in Fig.~\ref{fig:figures/conc_r1_relation}.
We fit both $R_1 (c)$ and $c(R_1)$ with third-order polynomial functions:
\begin{align}\label{eq:fitting} \log R_1 &= a_1(\log c)^3 + a_2(\log c)^2 +
    a_3\log c + a_4, \\
    \log c &= b_1(\log R_1)^3 + b_2(\log R_1)^2 + b_3\log R_1 + b_4,\nonumber \\
    a_1 &= 0.0198,~~a_2=-0.086, ~~a_3 = -0.090, ~~a_4 = -0.230,\nonumber\\
    b_1 &= -34.01,~~b_2=-43.91, ~~b_3 = -23.49, ~~b_4 = -3.48.\nonumber
\end{align}
The bottom panels of Fig.~\ref{fig:figures/conc_r1_relation} show the
fractional difference between the relation in equation~(\ref{eq:r1_c}) and the
fitting formula of equation~(\ref{eq:fitting}). The fractional deviation is
$\lesssim 0.1\%$ for the $R_1-c$ relation and $\lesssim 3\%$ for the $c-R_1$
relation. We note that the relation between $R_1$ and $c$ depends neither on
cosmology nor on the threshold density chosen to define dark matter halos.

\subsection{The NFW profile fitting method}%
\label{sub:the_profile_fitting_method}

The halo concentration can also be estimated by fitting the density
distribution with an NFW profile
\citep[e.g.][]{bhattacharyaDarkMatterHalo2013}. One can start with the
cumulative mass distribution for an NFW halo:
\begin{equation}
    M(<r) = \frac{m(cr/r_{\rm vir})}{m(c)}M_{\rm vir}\,,
\end{equation}
where
\begin{equation}
    m(x) = \ln(1 + x) - x/(1+ x)\,.
\end{equation}
The optimal concentration can be found by minimizing the $\chi^2$ defined as
\begin{equation}
    \chi^2 = \sum_i\frac{(M_i^{\rm sim} - M_i)^2}{(M_i^{\rm sim})^2/n_i},
\end{equation}
where $M_i = M(<r_i) - M(<r_{i-1})$ is the mass within the $i$-th radial bin
according to the NFW profile, $M_i^{\rm sim}$ is the total mass of particles in
the same radial bin for the simulated halo, and $n_i$ is the number of
particles in that bin. Here we take 20 equally spaced radial bins from
$0.01r_{\rm vir}$ to $r_{\rm vir}$ on the logarithmic scale. Clearly, the
result of the NFW profile fitting method is subject to the choice of binning.
In Appendix~\ref{sec:impact_of_binning_on_nfw_profile_fitting}, we test the
performance with three different binning strategies and adopt the best one here
to compare with the other two methods.

\subsection{The $V_{\rm max}/V_{\rm vir}$ method}%
\label{sub:the_v__rm_max_v__rm_vir_method}

For NFW halos, the concentration parameter is also related to the ratio between
the maximum circular velocity and the virial velocity,
\begin{equation} \label{eq:vmax} \frac{V_{\rm max}}{V_{\rm vir}} =
    \frac{{\tt{MAX}}(V_{\rm circ}(r))}{V_{\rm circ}(r_{\rm vir})}\,,
\end{equation}
where $V_{\rm circ}(r) = \sqrt{GM(<r)/r}$, and the relation is
\begin{equation} \label{eq:vmax_c} \frac{V_{\rm max}}{V_{\rm vir}} =
    \left[\frac{0.216 c}{\ln(1 + c) - c/(1+ c)}\right]^{1/2}
\end{equation}
\citep[e.g.][]{klypinResolvingStructureCold2001}. Note that this relation is
only applicable for $c \gtrsim 2.16$ since $V_{\rm max}\geq V_{\rm vir}$ by
definition.

\section{Testing the performance of the halo concentration estimators}%
\label{sec:performance}

\subsection{Tests on ideal NFW profile}%
\label{sub:tests_on_ideal_nfw_profile}

We first test the performance of the three concentration estimation methods on
ideal NFW halos generated from the \textsc{HaloFactory} package (see
appendix~\ref{sec:halo_factory}). The results are presented in
Fig.~\ref{fig:figures/conc_ideal_test}. The four columns are for four different
input halo concentrations, from $c=1$ to $c=20$, and the three rows present the
results for three different methods. In each panel, the red solid line shows
the median fractional deviation of the concentration parameter estimated from
10,000 halo realizations as a function of particle number, and the magenta
dashed lines and the cyan dotted lines show the $16^{\rm th}-84^{\rm th}$ and
$2.5^{\rm th}-97.5^{\rm th}$ percentile ranges, respectively.

Firstly, when the particle number is sufficiently large ($\gtrsim 10^4$), all
three methods perform equally well and the fractional deviation of the halo
concentration estimation for 95\% of halo realizations is within $\pm 5\%$.
Secondly, when the particle number decreases to a few hundred, the NFW profile
fitting method tends to underestimate halo concentration by $\approx 10\%$. We
note that this result is subject to the choice of binning (see
Appendix~\ref{sec:impact_of_binning_on_nfw_profile_fitting}). The $V_{\rm
max}/V_{\rm vir}$ method tends to overestimate the halo concentration by
$\approx 30\%$, which was already noted in previous studies
\citep[see][]{poveda-ruizQuantifyingControllingBiases2016}. In contrast, the
fractional deviation of the median value for our $R_1$ method is less than
$0.5\%$. Thirdly, the distribution of the estimated concentration broadens with
decreasing particle numbers. When only 100 particles are used, the width of the
$16^{\rm th}-84^{\rm th}$ percentiles is about $0.76c-1.03c$ for the NFW
fitting method, $0.91c-1.06c$ for the $V_{\rm max}/V_{\rm vir}$ method, and
$0.63c-0.91c$ for our $R_1$ method. Therefore, the $R_1$ method also yields the
smallest variance among all three methods when halos are poorly sampled. We
also note that the $V_{\rm max}/V_{\rm vir}$ method is not applicable to halos
with $c\lesssim 2.16$. In addition,
Appendix~\ref{sec:other_methods_to_estimate_halo_concentration} presents the
performance of three other concentration estimation methods. Our test results
show that their performances are poorer than the $R_1$ method, even though some
of them are more difficult to obtain from simulation data.

Finally, Appendix~\ref{sec:uncertainties} shows the distributions of the
logarithmic deviation of halo concentration estimated with our $R_1$ method.
These distributions can be described by Gaussian functions, and the scatter
decreases with increasing particle number and decreasing input concentration. A
fitting function is provided to describe the dependence of the scatter on the
particle number and the input concentration (see
equation~(\ref{eq:sigma_fitting})).

\subsection{Impact of resolution}%
\label{sub:impact_of_resolution}

\begin{figure}
    \centering
    \includegraphics[width=\linewidth]{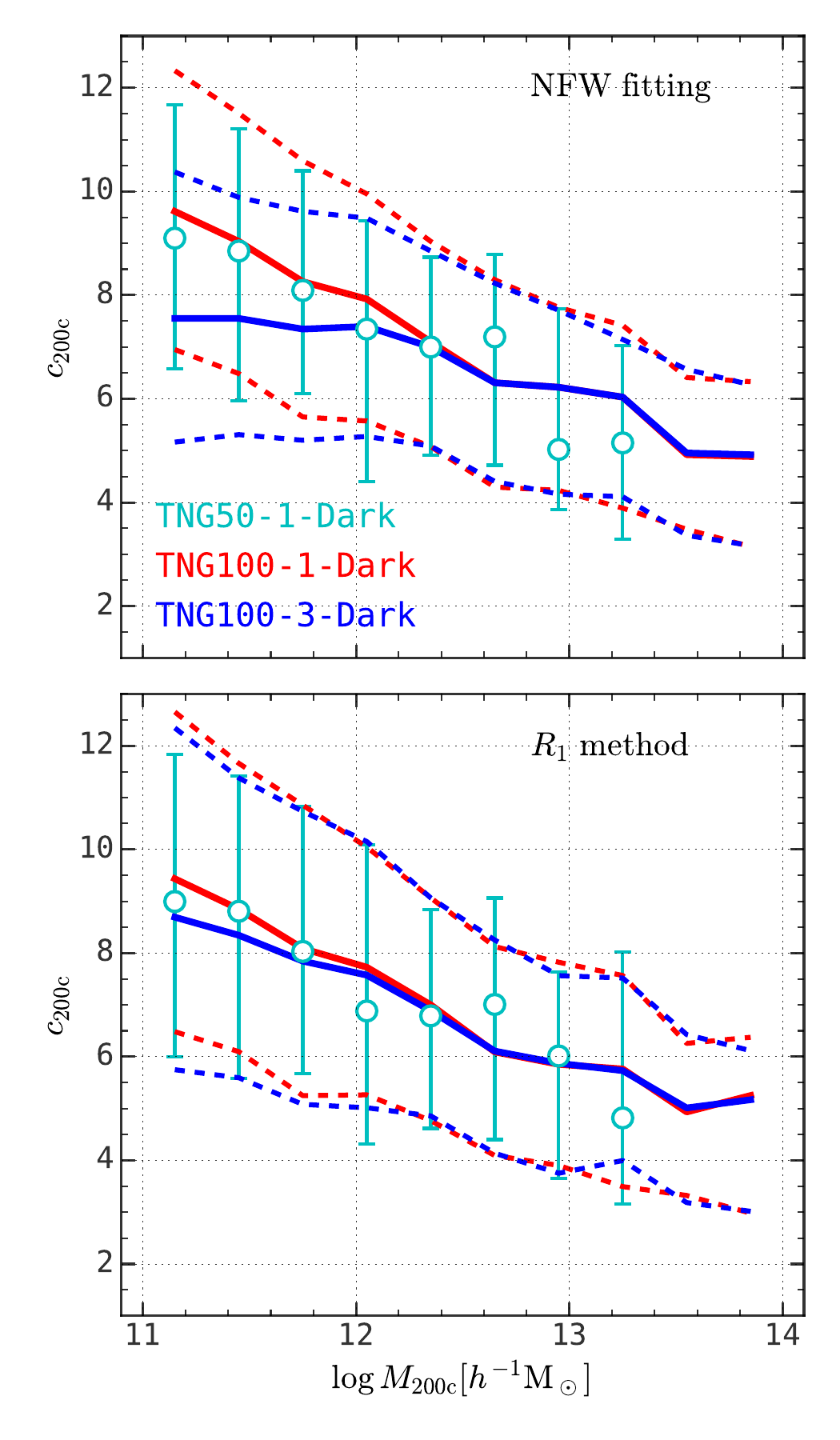}
    \caption{
        The mass-concentration relation for TNG50-1-Dark (cyan error bars),
        TNG100-1-Dark (red lines), and TNG100-3-Dark (blue lines), where the
        $16^{\rm th}-50^{\rm th}-84^{\rm th}$ percentiles are presented. The
        upper panel shows the result obtained with the NFW profile fitting
        method, and the lower panel shows the result obtained with our $R_1$
        method. The NFW profile fitting method underestimates halo
        concentration for low-resolution simulations, while this effect is
        marginal for our $R_1$ method.
    }%
    \label{fig:figures/mass_concentration_relation_tng}
\end{figure}

\begin{figure*}
    \centering
    \includegraphics[width=\linewidth]{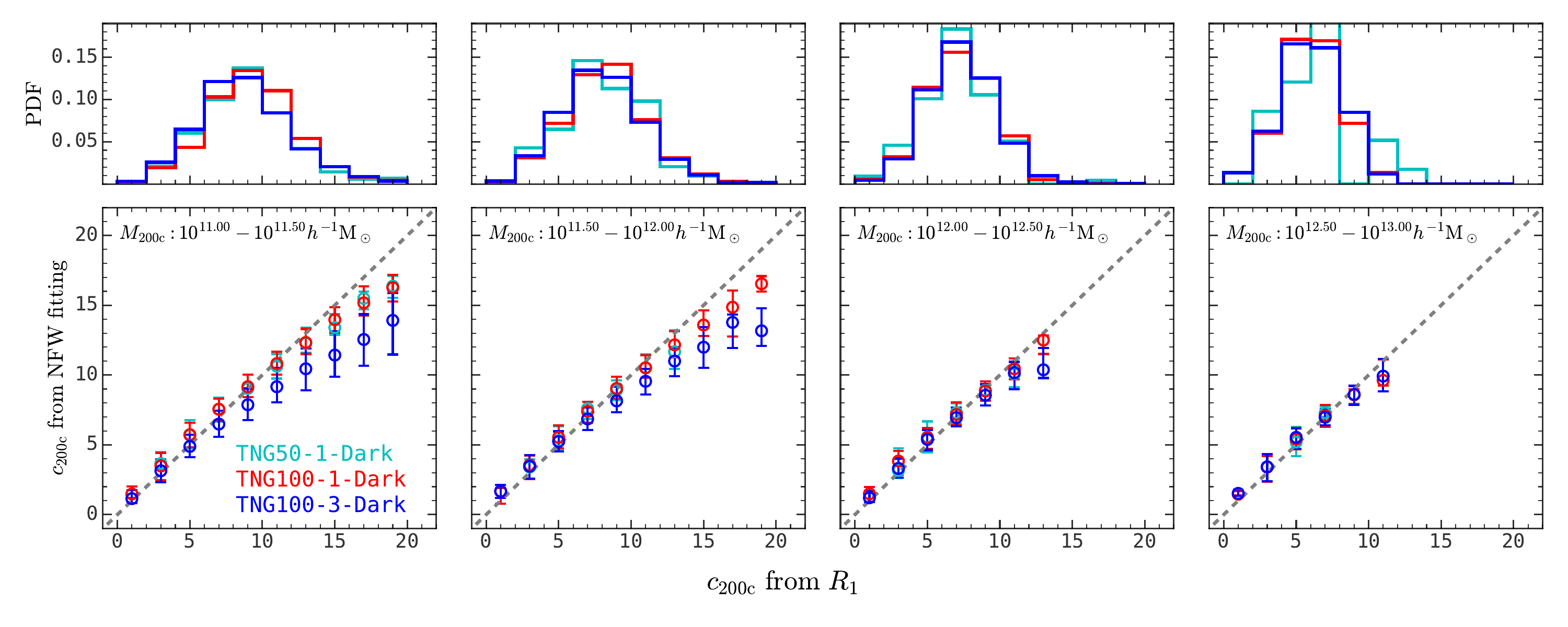}
    \caption{
        Top panels: The probability distribution function of halo concentration
        estimated with our $R_1$ method in three TNG-Dark simulations. Bottom
        panels: Comparison of halo concentration estimated from the NFW profile
        fitting method and our $R_1$ method in different halo mass bins. When
        combined with Fig.~\ref{fig:figures/mass_concentration_relation_tng},
        this figure demonstrates that, compared with our $R_1$ method, the NFW
        fitting method underestimates halo concentration for halos sampled with
        small numbers ($\lesssim 300$) of particles, especially for
        high-concentration halos.
    }%
    \label{fig:figures/conc_compare_r1_fitting_tng100_1_3}
\end{figure*}

We have already shown that our $R_1$ method outperforms the other two methods
in halo concentration estimation using ideal NFW halos. However, low-mass halos
in realistic $N$-body simulations are not only poorly sampled, but also subject
to the force softening used to avoid unphysical gravity when two particles are
too close to each other \citep[e.g.][]{powerInnerStructureLCDM2003,
ludlowNumericalConvergenceSimulations2019}. In addition, halos in simulations
are not spherically symmetric and are not perfectly relaxed
\citep[e.g.][]{jingDensityProfileEquilibrium2000,
jingTriaxialModelingHalo2002}. While it is not a priori clear what the true
concentration is for halos in simulations, we can investigate which method
gives the best convergence with the numerical resolution.

Here we use three simulations from the IllustrisTNG suite, which are
TNG50-1-Dark, TNG100-1-Dark, and TNG100-3-Dark, to test the impact of numerical
resolution on the estimation of halo concentration. Note again that the latter
two simulations use identical initial conditions and simulation code, but
different mass resolutions and gravitational softening lengths (see
table~\ref{tab:simulation}).
Fig.~\ref{fig:figures/mass_concentration_relation_tng} shows the
mass-concentration relation obtained from these three simulations, and the top
and bottom panels show results obtained from the NFW profile fitting method and
our $R_1$ method, respectively. Firstly, both methods produce nearly identical
median mass-concentration relations and the $16^{\rm th}-84^{\rm th}$
percentiles in TNG100-1-Dark, whose particle mass is about $6.0\times
10^6h^{-1}\rm M_\odot$. Secondly, the NFW profile fitting method underestimates
the concentration of $10^{11}h^{-1}\rm M_\odot$ halos ($\lesssim 300$ particles) by
$\gtrsim 20\%$ in TNG100-3-Dark, whose particle mass is about $3.8\times
10^8h^{-1}\rm M_\odot$. In contrast, the $R_1$ method yields nearly identical
mass-concentration relations across the entire mass range in these two
simulations, and the fractional deviation for low-mass halos is $\lesssim 8\%$
between the two simulations. Note that a $10^{11}h^{-1}\rm M_\odot$
halo in TNG100-3-Dark is represented by only $\lesssim 300$ particles. Finally,
the mass-concentration relations obtained from TNG50-1-Dark by the two methods
are similar to each other, and to those obtained from TNG100-1-Dark. The
discrepancy at the massive end owes to cosmic variance, since the two
simulations have different box sizes and initial conditions. In
Appendix~\ref{sec:c500c}, we show that the concentration parameter $c_{\rm
200c}$ can also be obtained from $R_1$ with integrating only to $r_{\rm 500c}$,
which is commonly used in observation.

Fig.~\ref{fig:figures/conc_compare_r1_fitting_tng100_1_3} compares the halo
concentration estimated with the NFW profile fitting method and our $R_1$
method in the three TNG-Dark simulations, where open circles and error bars
show the median and the $16^{\rm th}-84^{\rm th}$ percentiles, respectively.
Firstly, both methods yield similar concentrations for massive halos and
low-concentration low-mass halos. Secondly, the NFW fitting method produces
lower concentrations than our $R_1$ method for high-concentration low-mass
halos, and the discrepancy is larger in lower-resolution simulations. Combined
with the results in Fig.~\ref{fig:figures/mass_concentration_relation_tng}, we
infer that the NFW profile fitting method tends to underestimate the
concentration of high-concentration low-mass halos in low-resolution
simulations for two reasons. The first one is that the NFW profile fitting
method tends to underestimate halo concentration for poorly-sampled halos, as
shown in Fig.~\ref{fig:figures/conc_ideal_test}, but this effect becomes
marginal once more than a few thousand particles are sampled. The second reason
is that, for a given simulation volume, the force softening length is larger in
lower resolution runs, and so is a larger fraction of the virial radius in
lower mass halos, and therefore has a large impact on the central mass profile.
This will consequently cause the underestimation of halo concentration in the
NFW profile fitting method. A common strategy to tackle this problem is to
exclude particles below the convergence radius during the fitting, where the
convergence radius is defined such that the two-body dynamical relaxation
timescale of the particles within this radius is comparable to the age of the
universe \citep{powerInnerStructureLCDM2003, duffyDarkMatterHalo2008a,
correaAccretionHistoryDark2015}. However, the convergence radius is about
$0.1r_{\rm 200c}$ for halos with a few hundred particles
\citep{ludlowNumericalConvergenceSimulations2019}, and excluding particles
within this radius will cause systematic underestimations of the concentration
parameter by $\approx 20\%-50\%$ for halos with $c\approx 10-20$ for the NFW
fitting method (see
Appendix~\ref{sec:impact_of_binning_on_nfw_profile_fitting}). In contrast, the
$R_1$ method is less affected by the inclusion, since it gives more weight to
the outer region of dark matter halos.

Finally, there is still a noticeable discrepancy between these two methods for
high-concentration halos with $M_{\rm 200c}\approx 10^{11}h^{-1}\rm M_\odot$ in
TNG100-1-Dark and TNG50-1-Dark, where a $10^{11}h^{-1}\rm M_\odot$ halo is well
represented by $\gtrsim 2.7\times 10^5$ particles. In
Appendix~\ref{sec:outlier}, we find that these halos deviate from the NFW
profile due to the stripping of mass in the outskirt, and the mass distribution
recovered from both concentrations matches the data equally well, despite the
$\gtrsim 10\%$ systematics in the values of the estimated concentration.
Nevertheless, these halos constitute only a small portion of all halos in the
given mass bin, as one can see from the histogram in the top panels of
Fig.~\ref{fig:figures/conc_compare_r1_fitting_tng100_1_3}.

\section{The mass-concentration relation in the ELUCID simulation}
\label{sec:application_to_elucid}

\begin{figure*}
    \centering
    \includegraphics[width=0.9\linewidth]{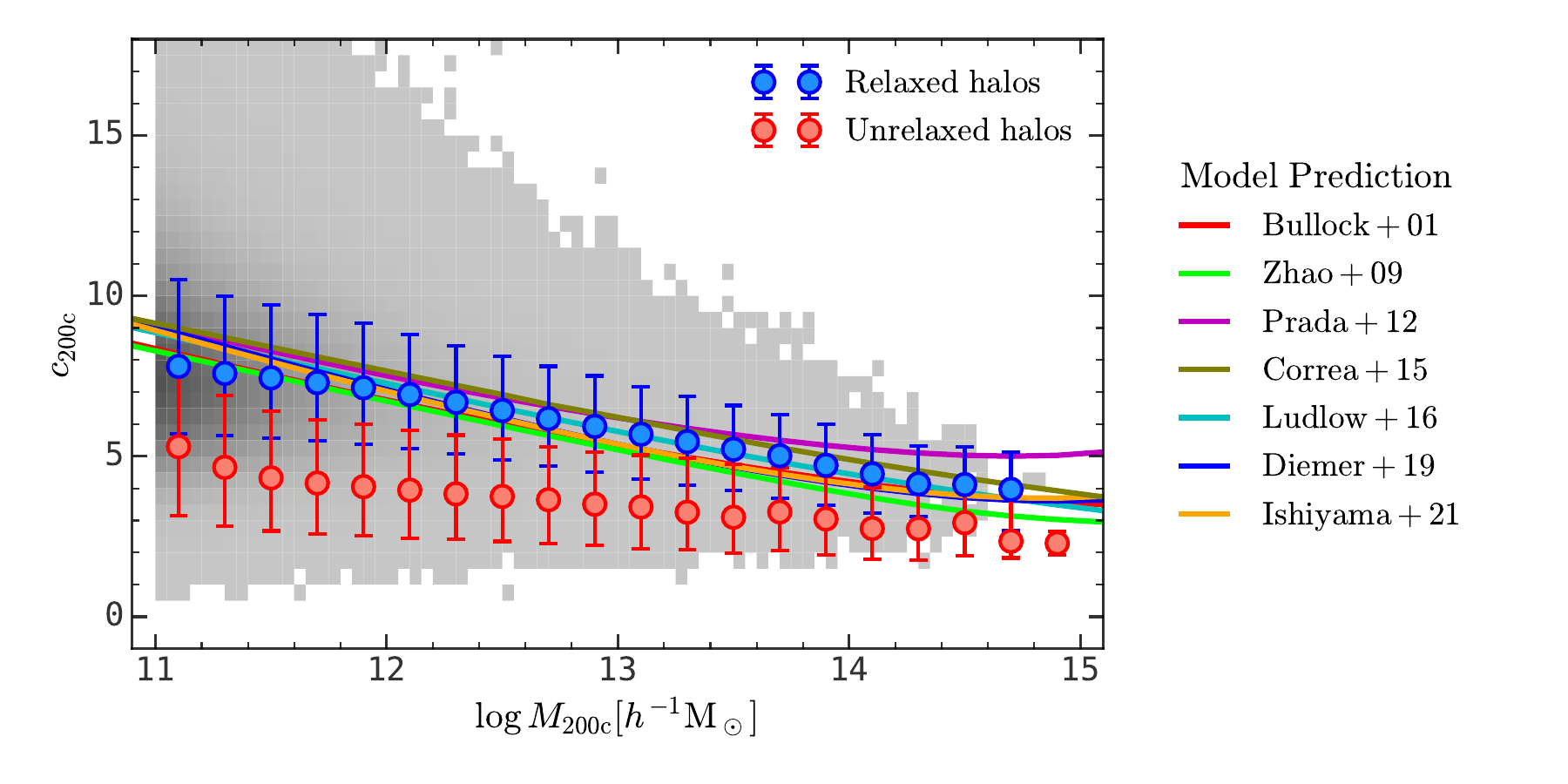}
    \caption{
        The mass-concentration relation in the ELUCID simulation for relaxed
        (blue error bars) and unrelaxed (red error bars) halos, and the
        $16^{\rm th}-84^{\rm th}$ percentiles. The gray color scale encodes the
        number density of dark matter halos. The criterion for separating relaxed
        and unrelaxed halos is shown in equation~(\ref{eq:relax_halo}). The solid
        lines are the predictions of different semi-analytical models:
        Bullock+01 \citep{bullockProfilesDarkHaloes2001}, Zhao+09
        \citep{zhaoACCURATEUNIVERSALMODELS2009}, Prada+12
        \citep{pradaHaloConcentrationsStandard2012}, Correa+15
        \citep{correaAccretionHistoryDark2015}, Ludlow+16
        \citep{ludlowMassconcentrationredshiftRelationCold2016}, Diemer+19
        \citep{diemerAccuratePhysicalModel2019}, and Ishiyama+21
        \citep{ishiyamaUchuuSimulationsData2021}.
    }%
    \label{fig:figures/mass_concentration_relation}
\end{figure*}

\begin{figure*}
    \centering
    \includegraphics[width=1\linewidth]{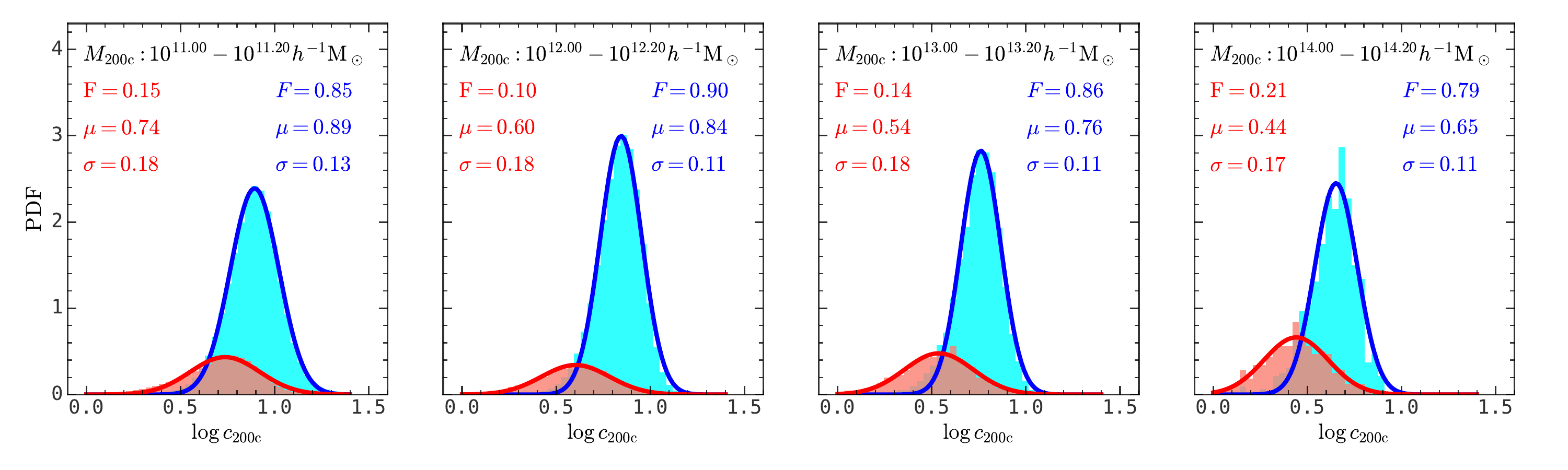}
    \caption{
        The distribution of the logarithmic halo concentration in four mass
        bins for relaxed (blue) and unrelaxed (red) halos, where the criterion
        to separate these two populations is shown in
        equation~(\ref{eq:relax_halo}). The distribution of both halo populations
        are fitted with a Gaussian function as shown in solid lines. The
        best-fitting parameters are shown on the panel.
    }%
    \label{fig:figures/mass_conc_scatter_elucid}
\end{figure*}

\begin{figure*}
    \centering
    \includegraphics[width=0.9\linewidth]{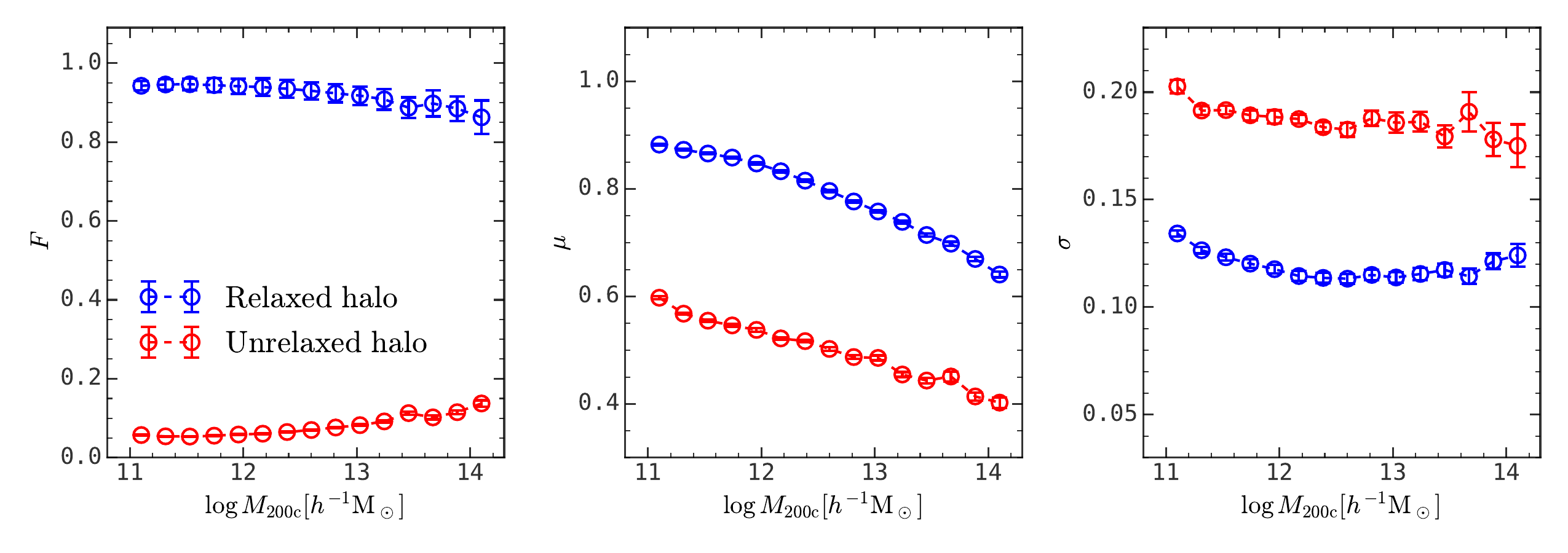}
    \caption{
        The halo mass dependence of the fitting parameters for relaxed (blue)
        and unrelaxed (red) halos, where the left panel shows the halo fraction
        $F$, the middle panel the mean $\log c$, and the right panel the
        standard deviation of $\log c$.
    }%
    \label{fig:figures/mass_conc_scatter_depen_elucid}
\end{figure*}

It has been shown in \S\,\ref{sec:performance} that our $R_1$ method
outperforms the conventional method in the halo concentration estimation on
both ideal NFW halos and realistic halos in $N$-body simulations, and it can
give unbiased estimation of the concentration parameter for halos with more
than 200 particles. For this reason, we apply the $R_1$ method to the ELUCID
simulation and infer the mass-concentration relation for halos with $11\lesssim
\log(M_{\mathrm 200c}/[h^{-1}{\rm M_\odot}])\lesssim 15$. Notably, a
$10^{11}h^{-1}\rm M_\odot$ halo is only represented by about $300$ particles
in ELUCID.

Fig.~\ref{fig:figures/mass_concentration_relation} shows the median
mass-concentration relation in ELUCID, as well as the $16^{\rm th}-84^{\rm th}$
percentiles. Here relaxed and unrelaxed halos are separated according to the
criterion in \citet{netoStatisticsCDMHalo2007}, which is
\begin{align}
    \label{eq:relax_halo} {\rm Relaxed~halos:}& \Delta < 0.07r_{\rm 200c}\\
    {\rm Unrelaxed~halos:}& \Delta > 0.07r_{\rm 200c}\nonumber\\
                          &\Delta=\|\mathbf r_{\rm min-pot} - \mathbf r_{\rm
                          com}\|\nonumber
\end{align}

where $\mathbf r_{\rm min-pot}$ is the position of the particle with the
minimal gravitational potential, and $\mathbf r_{\rm com}$ is the center of
mass of all dark matter particles within $r_{\rm 200c}$. Note that
\citet{netoStatisticsCDMHalo2007} uses two additional conditions to select
halos in equilibrium. They require that the mass fraction in substructures is
lower than a threshold value and that the ratio between the kinetic energy and
the potential energy is lower than a threshold. Here we use only the criterion
in equation~(\ref{eq:relax_halo}), for three reasons. Firstly, as shown in
\citet{netoStatisticsCDMHalo2007}, equation~(\ref{eq:relax_halo}) alone can
select most of the halos in equilibrium (see their Figure~2). Secondly,
equation~(\ref{eq:relax_halo}) is the simplest criterion to implement in N-body
simulations, whereas the other two criteria require either identifying
substructures or calculating the gravitational potential for each particle.
Thirdly, the other two criteria suffer from some ambiguities. For instance, the
substructure mass fraction is subject to the substructure finder used
\citep[e.g.][]{vandenboschStatisticsDarkMatter2016} and to the resolution of
the simulation \citep[e.g.][]{vandenboschDisruptionDarkMatter2018}. Besides,
the exact value of the virial ratio for selecting halos in equilibrium is still
under debate, as many argued that the surface pressure and even the
non-spherical shape of halos should be taken into account
\citep[e.g.][]{davisVirializationHighredshiftDark2011,
klypinMultiDarkSimulationsStory2016}. Here one can see that the concentration
parameter decreases from $\approx 8$ to $\approx 4$ with increasing mass for
relaxed halos, and from $\approx 4$ to $\approx 2$ for unrelaxed ones. It has
already been noted in previous studies that unrelaxed halos exhibit lower
concentration than relaxed ones
\citep[e.g.][]{jingDensityProfileEquilibrium2000, netoStatisticsCDMHalo2007,
duffyDarkMatterHalo2008a, childHaloProfilesConcentrationMass2018}. A detailed
analysis in \citet{wangConcentrationsDarkHaloes2020} reveals that a sudden
halo-halo merger event will reduce the concentration dramatically, and the
concentration parameter will gradually increase during the subsequent secular
evolution. For comparison, the solid lines show the mass-concentration
relations given by seven different semi-analytical models with the same
cosmology and halo definition\footnote{All these models are implemented in the
    \textsc{Colossus} package \citep{colossus}, except Zhao+09
    (\url{http://202.127.29.4/dhzhao/mandc.html}) and Correa+15
(\url{https://www.camilacorrea.com/code/commah/}).}. Our results are broadly
consistent with these models.

In addition to the median mass-concentration relation, the distribution of
concentration at given halo masses also carries important information.
Fig.~\ref{fig:figures/mass_conc_scatter_elucid} shows the distribution of the
logarithmic halo concentration, $\log c$, for relaxed and unrelaxed halos in
four narrow halo mass bins. For each halo population in a given mass bin, we
fit the distribution of $\log c$ to a Gaussian function. Each distribution is
thus described by three parameters: $F$ as the fraction of the target halo
population among all halos in the same mass bin, $\mu$ as the mean of the
Gaussian function, and $\sigma$ as the standard deviation. The fitting
functions are shown in blue and red solid lines in
Fig.~\ref{fig:figures/mass_conc_scatter_elucid} for relaxed and unrelaxed
halos, respectively. One can see that the Gaussian model describes the
distribution quite well.

Fig.~\ref{fig:figures/mass_conc_scatter_depen_elucid} shows the halo mass
dependence of these fitting parameters. First of all, the unrelaxed halos only
amounts to about $5\%$ of all halos with $M_{\rm 200c}\approx 10^{11}h^{-1}\rm
M_\odot$, and this fraction increases to about $15\%$ for $10^{14}h^{-1}\rm
M_\odot$ halos. The positive correlation between the unrelaxed halo fraction
and halo mass is expected, since the halo merger rate is positively correlated
to halo mass \citep[e.g.][]{fakhouriNearlyUniversalMerger2008}. Secondly, the
mean logarithmic concentration declines with increasing halo mass for both
relaxed and unrelaxed halos, with a constant gap of about $0.28$ dex. Finally,
the scatter in the distribution of $\log c$ for relaxed and unrelaxed halos are
about $0.12$ and $0.19$ dex, respectively, with a weak dependence on halo mass.

\section{Other applications of $R_1$}%
\label{sec:other_applications_of_r_1_}

\subsection{Estimating $V_{\rm max}$ from $R_1$}%
\label{sub:estimating_v__rm_max_from_r_1_}

\begin{figure*}
    \centering
    \includegraphics[width=1\linewidth]{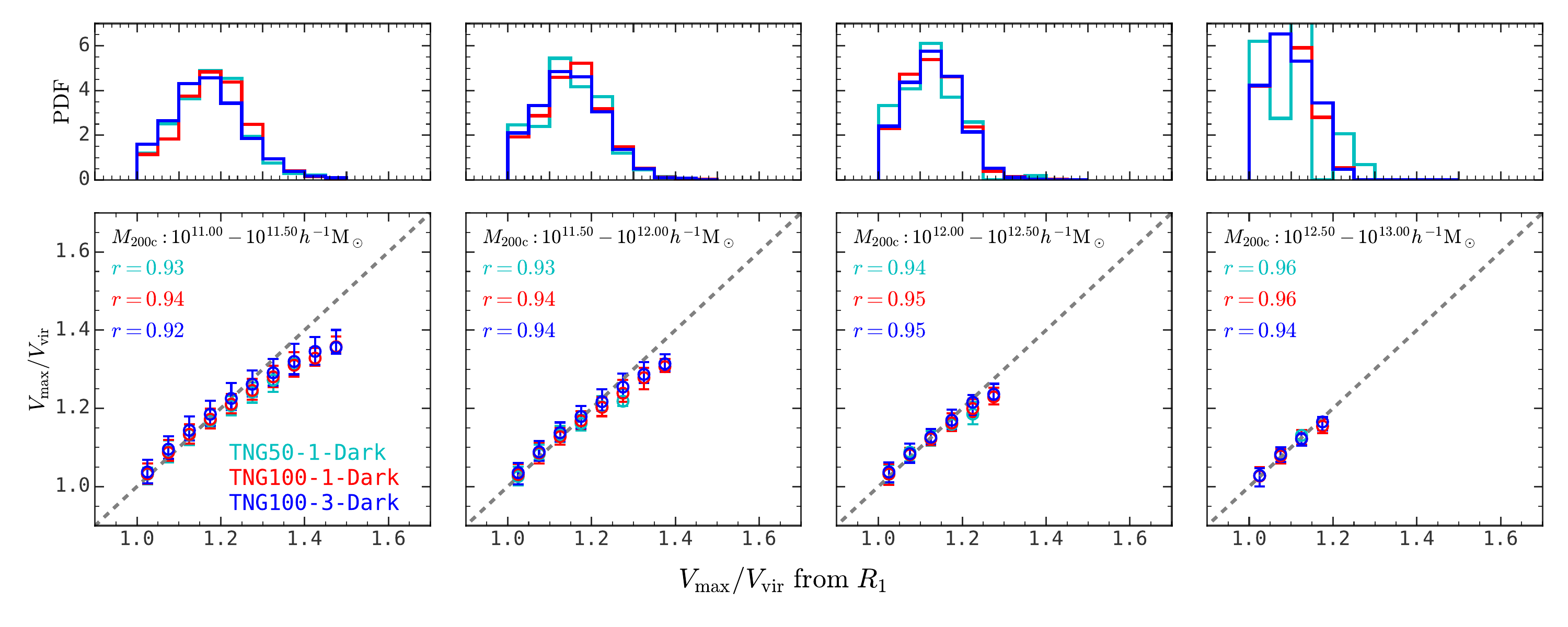}
    \caption{
        Top panels: The probability distribution of $V_{\rm max}/V_{\rm vir}$
        estimated from $R_1$ in three TNG-Dark simulations. Bottom panels:
        Comparison of $V_{\rm max}/V_{\rm vir}$ calculated from
        equation~(\ref{eq:vmax}) and from our $R_1$ method in three TNG-Dark
        simulations. Spearman's rank correlation coefficients are labeled on
        the bottom panels.
    }%
    \label{fig:figures/vmax_compare_r1_fitting_tng100_1_3}
\end{figure*}

The maximum circular velocity, $V_{\rm max}$, is not only a proxy of halo
concentration, but also a commonly-adopted quantity to connect galaxies with
their dark matter halos \citep{reddickConnectionGalaxiesDark2013,
mattheeOriginScatterStellar2017, zehaviProspectUsingMaximum2019}. It is thus
important to be able to obtain $V_{\rm max}$ efficiently and robustly for a
large sample of simulated halos in order to investigate the galaxy-halo
connection using large cosmological simulations. To this end, we derive $V_{\rm
max}$ from $R_1$ according to equations~(\ref{eq:r1_c}) and \ref{eq:vmax_c}.
Fig.~\ref{fig:figures/vmax_compare_r1_fitting_tng100_1_3} compares the $V_{\rm
max}/V_{\rm vir}$ calculated from equation~(\ref{eq:vmax}) and derived from
$R_1$, where one can see they match quite well. We note that there is a small
discrepancy for low-mass halos with high $V_{\rm max}/V_{\rm vir}$, which has
the same origin as the discrepancy for low-mass halos with high concentrations
shown in Fig.~\ref{fig:figures/conc_compare_r1_fitting_tng100_1_3} (see also
Appendix~\ref{sec:outlier}). Nevertheless, these halos only account for a small
portion of all halos at the given halo mass bin, as shown in the top panels of
Fig.~\ref{fig:figures/vmax_compare_r1_fitting_tng100_1_3}. The relative rank is
well preserved, as indicated by high Spearman's rank correlation coefficients
($\gtrsim 0.92$).

It should be noted that $R_1$ is defined only for main halos\footnote{In
    principle the $R_1$ method can also be used for stripped satellite subhalos
    provided the core survives. The integral in equation~(\ref{eq:r1}) should
    then be stopped before the virial radius at some $R_\Delta$ with $\Delta >
200$ (see Appendix~\ref{sec:c500c}).}. For a satellite subhalo contained in
a host halo, one can trace its main-branch progenitor to the snapshot prior
to the infall into its host halo and calculate its $R_1$ to derive $V_{\rm
max}$. This is similar to the calculation of $V_{\rm peak}$, which is the
peak value of $V_{\rm max}$ on the main branch and serves as a better proxy
in subhalo abundance matching than $V_{\rm max}$
\citep{reddickConnectionGalaxiesDark2013}. However, it is unclear whether
or not environmental effects prior to the infall of halos can break the
relation between $V_{\rm max}$ and $R_1$. To test the validity of the $R_1$
method for subhalos, we compare results between pre-infall halos at a given
redshift, defined as halos that will become subhalos in the subsequent
snapshot, and the results are presented in
Appendix~\ref{sec:_v__rm_max_estimation_for_pre_infall_halos}. There one
can see that the $V_{\rm max}$ - $R_1$ relation does not depend on whether
or not halos are soon falling into other halos to become a satellite,
indicating that the $R_1$ method can also be used to estimate $V_{\rm
peak}$ for subhalos.

\subsection{Estimating the Einasto concentration from $R_1$}%
\label{sub:einasto_concentration}

It has been suggested that the radial density distribution of dark matter halos
in $N$-body simulations is better fitted with a three-parameter Einasto profile
\citep{navarroInnerStructureLCDM2004, gaoRedshiftDependenceStructure2008,
wangUniversalStructureDark2020}, which has the form
\begin{equation}
    \label{eq:einasto}
    \rho(r) = \rho_{-2}\exp\left\{
        -\frac{2}{\alpha}\left[\left(\frac{r}{r_{-2}}\right)^\alpha -
    1\right]\right\}\,,
\end{equation}
where $\rho_{-2}$, $\alpha$, and $r_{-2}$ are free parameters.
\citet{gaoRedshiftDependenceStructure2008} found that there is a universal
relationship between $\alpha$ and the peak height $\nu$ given by
\begin{align}
    \alpha = 0.155 + 0.0095v^2, ~\nu = \delta_{\rm crit}(z)/\sigma(M_{\rm vir},
    \label{eq:alpha} z)
\end{align}
where the peak height $\nu$ is defined as the ratio between the critical
overdensity $\delta_{\rm crit}(z)$ for collapse at redshift $z$ and the linear
rms fluctuation at $z$ within spheres containing mass $M_{\rm vir}$. We note
that the value of $\nu$ is determined by redshift and halo mass for a given
cosmology. The typical value of $\alpha$ is between 0.15 and 0.3. The
concentration parameter for the Einasto profile is defined as
\begin{equation}
    c_{\rm e} \equiv r_{\rm vir} / r_{-2}\,.
\end{equation}
Therefore, at a given redshift of $z$, the halo mass $M_{\rm vir}$ and the
Einasto concentration $c_{\rm e}$ together determine the halo density profile,
with the parameter $\alpha$ determined by equation~(\ref{eq:alpha}).

\begin{figure}
    \centering
    \includegraphics[width=\linewidth]{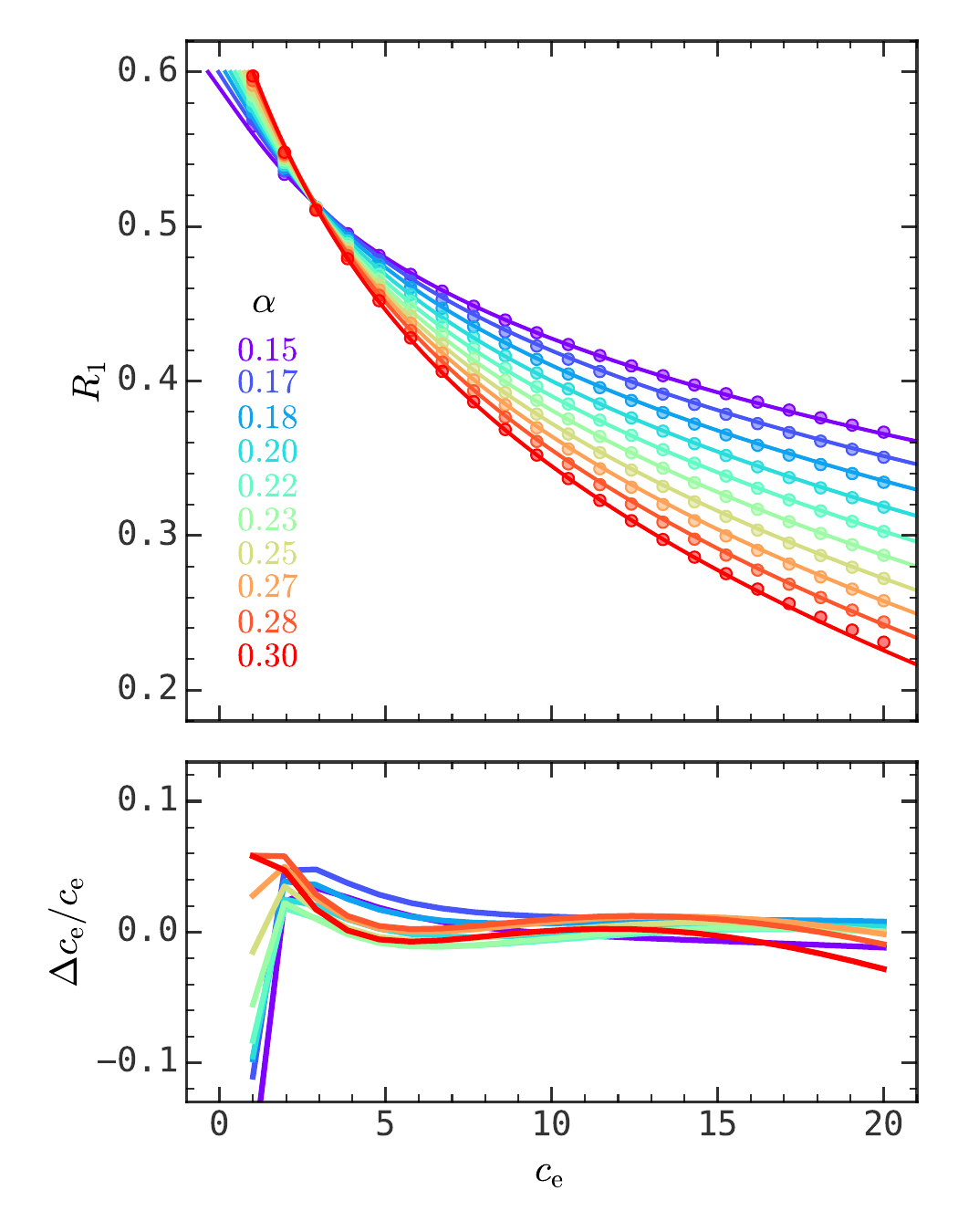}
    \caption{
        The relation between $R_1$ and the concentration parameter of the
        Einasto profile, $c_{\rm e}$, for different
        values of $\alpha$ (see equation~(\ref{eq:einasto})). The solid lines are
        the fitting function in equation~(\ref{eq:ce_fitting}).
    }%
    \label{fig:figures/einasto}
\end{figure}

Fig.~\ref{fig:figures/einasto} shows the relation between $R_1$ and the Einasto
concentration $c_{\rm e}$ for $0.15\leq\alpha\leq 0.3$ in circles. And the
solid lines are the fitting function,
\begin{align}\label{eq:ce_fitting} c_e &= d_1x^3 + d_2x^2 + d_3x + d_4\\
    x &= {R_1\over\alpha^{0.95}} + e_1\alpha^3 + e_2\alpha^2 + e_3\alpha +
    e_4\nonumber\\ d_1 &= -5.45,~~d_2 = 14.72,~~d_3= -18.70,~~d_4 =
    9.07\nonumber\\
    e_1 &= 191.32,~~e_2=-173.00,~~e_3= 57.78,~~e_4= -8.06\nonumber
\end{align}
The bottom panel shows the fractional residual, from which one can see that
this fitting function is accurate to $\lesssim 5\%$ for $c_{\rm e}\gtrsim 3$
and $\lesssim 10\%$ for $c_{\rm e}\lesssim 3$.

\section{Summary}%
\label{sec:summary}

Estimating the concentration parameter and related quantities of simulated dark
matter halos in large numerical $N$-body simulations is a critical step to
study halo structure and understand its relation to the halo assembly history
and to the properties of galaxies that form in them. A reliable and efficient
method is needed to estimate these quantities for large cosmological
simulations that include halos with a wide range of particle numbers. To this
end, we propose an efficient and robust method to estimate the halo
concentration and related quantities using the first moment of the density
distribution. Our main results are summarized as follows:
\begin{enumerate}

    \item We find that the first moment of the density distribution, defined as
        $R_1=\int_0^{r_{\rm vir}}4\pi r^3\rho(r)dr^3/M_{\rm vir}/r_{\rm vir}$,
        has a simple, monotonic relation with the halo concentration for NFW
        halos. A cubic polynomial function can describe this relation to
        $\lesssim 3\%$ accuracy (see Fig.~\ref{fig:figures/conc_r1_relation}).

    \item Testing on ideal NFW halos, we find that the NFW profile fitting
        method and the $V_{\rm max}/V_{\rm vir}$ method introduce $\approx
        10\%$ and $\approx 30\%$ systematics for halos with 100 particles. In
        contrast, the bias introduced by the $R_1$ method is smaller than
        $0.5\%$. The $R_1$ method yields the smallest variance among all the
        three methods.

    \item Testing on realistic halos in $N$-body simulations of different
        resolutions, we find that the NFW fitting method underestimates the
        concentration parameter of halos with $\lesssim 300$ particles by
        $\gtrsim 20\%$, due to the poor sampling and the large gravitational
        softening length. In contrast, such effects only introduce $\lesssim
        8\%$ systematics in the $R_1$ method (see
        Figs.~\ref{fig:figures/mass_concentration_relation_tng} and
        \ref{fig:figures/conc_compare_r1_fitting_tng100_1_3}).

    \item We apply the $R_1$ method to the ELUCID $N$-body simulation and
        obtain the mass-concentration relation across four orders of magnitude
        of halo mass, separately for relaxed and unrelaxed halos. We find that
        the distributions of the logarithmic concentration, $\log c$, for both
        populations can be described by a Gaussian function. We find that the
        fraction of unrelaxed halos ranges from $\approx 5\%$ to $\approx 15\%$
        from $10^{11}h^{-1}\rm M_\odot$ to $10^{14}h^{-1}\rm M_\odot$. The mean
        logarithmic concentration declines monotonically with halo mass for
        both relaxed and unrelaxed halos, and there is a constant difference of
        $\approx 0.28$ dex between unrelaxed halos of lower concentration and
        relaxed ones with higher concentration. The standard deviations of the
        logarithmic concentration for relaxed and unrelaxed halos are $\approx
        0.12$ dex and $\approx 0.19$ dex, respectively, with a weak dependence
        on halo mass. (see Figs.~\ref{fig:figures/mass_concentration_relation},
        \ref{fig:figures/mass_conc_scatter_elucid}, and
        \ref{fig:figures/mass_conc_scatter_depen_elucid}).

    \item The maximum circular velocity, $V_{\rm max}$, of simulated halos can be
        derived from $R_1$ efficiently.
        The $V_{\rm max}$ - $R_1$ relation is not affected by whether or not
        the halo in question is about to be accreted by another halo and to
        become a subhalo (see
        Fig.~\ref{fig:figures/vmax_compare_r1_fitting_tng100_1_3} and
        Appendix~\ref{sec:_v__rm_max_estimation_for_pre_infall_halos}).

    \item We find a fitting function for the relation between $R_1$ and the
        Einasto concentration $c_{\rm e}=r_{\rm vir}/r_{-2}$ with $0.15 \leq
        \alpha \leq 0.3$, and the fractional deviation is $\lesssim 5\%$ for
        $c\gtrsim 3$ and $\lesssim 10\%$ for $c\lesssim 3$ (see
        Fig.~\ref{fig:figures/einasto}).

\end{enumerate}

The concentration parameter and related structural quantities of dark matter
halos play an important role in the study of dark matter halos and the modeling
of the galaxy-halo connection. However, because of the uncertainty and tedium
in their estimations, many cosmological simulations run in large boxes with
relatively low resolutions avoid providing these quantities. The $R_1$ method
proposed here can fill the gap, as it provides an accurate proxy for the
concentration parameter for both NFW and Einasto halos. Its estimation is both
straightforward and efficient, thus suitable for large cosmological
simulations, such as MillenniumTNG
\citep{boseMillenniumTNGProjectLargescale2023} and FLAMINGO
\citep{schayeFLAMINGOProjectCosmological2023}. We suggest that this quantity be
provided in simulated halo catalogs along with other important halo properties.

\section*{Acknowledgements}

Kai Wang thanks Fangzhou Jiang for his helpful comments and suggestions. The
authors acknowledge the Tsinghua Astrophysics High-Performance Computing
platform at Tsinghua University for providing computational and data storage
resources that have contributed to the research results reported within this
paper. This work is supported by the National Science Foundation of China
(NSFC) Grant No. 12125301, 12192220, 12192222, and the science research grants
from the China Manned Space Project with NO. CMS-CSST-2021-A07. YC is supported
by China Postdoctoral Science Foundation Grant No. 2022TQ0329 and NSFC Grant
No. 12192224.

The computation in this work is supported by the HPC toolkits \textsc{HIPP}
\citep{hipp} and
\textsc{PyHIPP}\footnote{\url{https://github.com/ChenYangyao/pyhipp}},
\textsc{IPYTHON} \citep{perezIPythonSystemInteractive2007}, \textsc{MATPLOTLIB}
\citep{hunterMatplotlib2DGraphics2007}, \textsc{NUMPY}
\citep{vanderwaltNumPyArrayStructure2011}, \textsc{SCIPY}
\citep{virtanenSciPyFundamentalAlgorithms2020}, \textsc{ASTROPY}
\citep{astropy:2013, astropy:2018, astropy:2022}. This research made use of
NASA’s Astrophysics Data System for bibliographic information. The authors
thank ELUCID collaboration for making their data products publicly
available\footnote{\url{https://www.elucid-project.com/}}.

\section*{Data availability}

The data underlying this article will be shared on reasonable request to the
corresponding author.

\bibliographystyle{mnras}
\bibliography{bibtex.bib}

\appendix

\section{Generate dark matter halos with $\textsc{HaloFactory}$}%
\label{sec:halo_factory}

\begin{figure}
    \centering
    \includegraphics[width=1\linewidth]{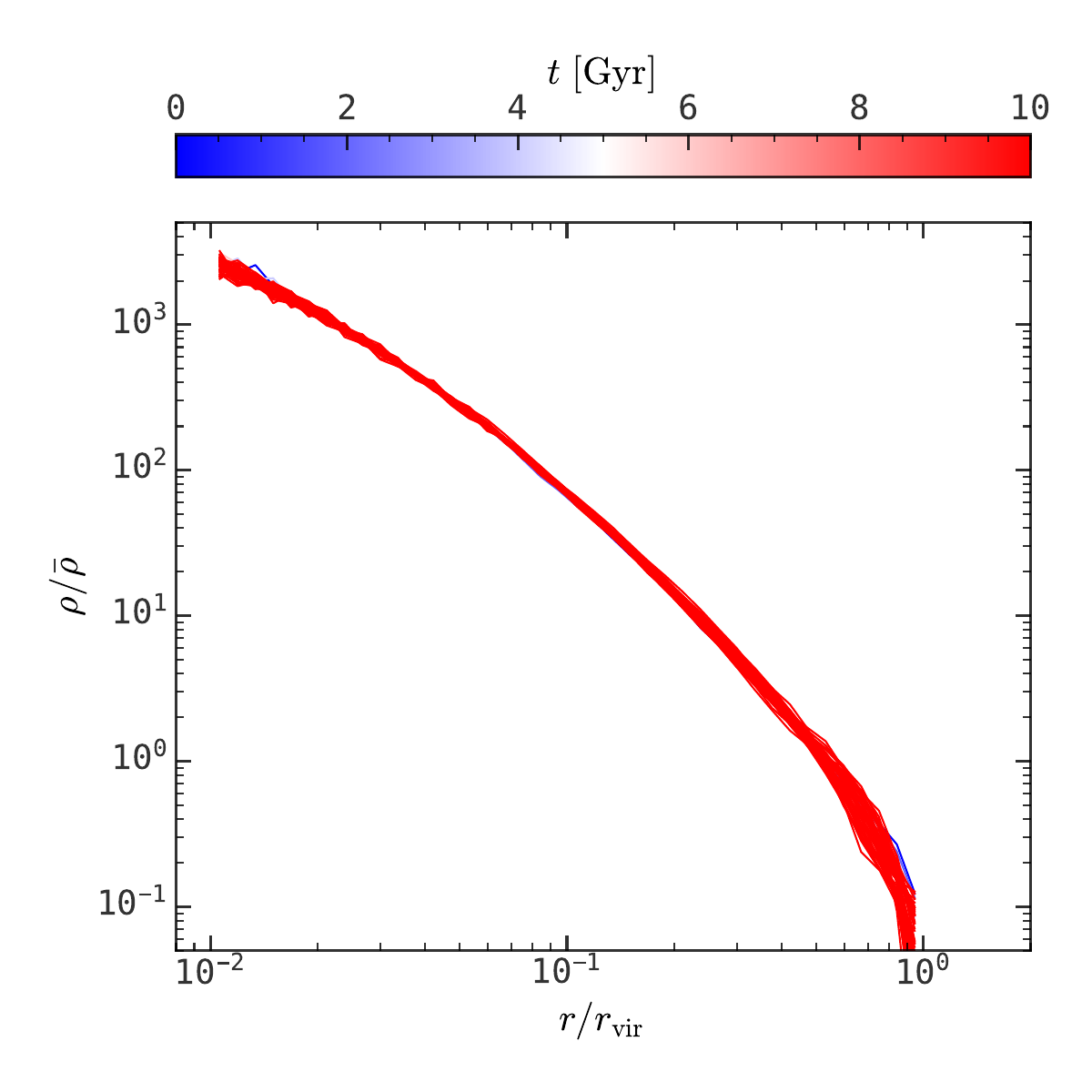}
    \caption{
        The time evolution of the density profile of a halo with 10,000
        particles generated from the \textsc{HaloFactory} package from 0 to 10
        Gyr.
    }%
    \label{fig:figures/halo_factory_evolution}
\end{figure}

Here we present an open-source Python package, \textsc{HaloFactory}\footnote{
\url{https://github.com/ChenYangyao/halofactory}}, which implements the
algorithms for generating the position and velocity of particles in mock halos.
These halos are assumed to be collisionless and spherically symmetric systems
in dynamical equilibrium, whose behavior can be described by the collisionless
Boltzmann equation \citep{2008gady.book.....B}. \textsc{HaloFactory} uses
Eddington's inversion method to sample the position and velocity of particles
from the exact phase-space distribution function, which is more accurate than
approximation methods that rely on the Jeans equations
\citep{2008gady.book.....B}. In addition, \textsc{HaloFactory} is rather
efficient due to the adoption of various numerical acceleration techniques.
Finally, \textsc{HaloFactory} can serve various purposes, such as providing
initial conditions for numerical simulations and serving as input for
halo-based galaxy models. Fig.~\ref{fig:figures/halo_factory_evolution} shows
the time evolution of the density profile of a halo generated from
\textsc{HaloFactory} with 10,000 particles and an initial concentration of 15.
Here one can see that the density profile has nearly no evolution on a time
scale of 10 Gyr, which indicates that the halos generated from
\textsc{HaloFactory} are already in equilibrium.

The implementation of \textsc{HaloFactory} relies on Eddington's formula. For
an equilibrium and spherical system given by the density distribution
$\rho(r)$, the particle energy distribution is given by
\begin{equation}
    f(E) = {1\over \sqrt{8}\pi^2}{\mathrm d\over \mathrm dE}\int_E^{E_{\rm
    T}}\mathrm dV{1\over V-E}{\mathrm d\rho(V)\over \mathrm dV},
\end{equation}
which is called Eddington’s formula \citep[see chapter 4
in][]{2008gady.book.....B}. Here $E$ is the particle energy and $V$ is the
potential energy. We also require the maximum energy of particles $E_{\rm T}$
equals to the potential energy at the truncation radius $r_{\rm T}$, which is a
free parameter, so that particles are restrained within $r_{\rm T}$. In
\textsc{HaloFactory}, the sampling of particle position and velocity are
proceeded as follows:
\begin{itemize}
    \item {\bf Sampling particle position:} We can integrate the radial density
        profile $\rho(r)$ to get the accumulative mass distribution, which is
        $M(<r) = \int_0^r 4\pi x^2\rho(x)dx$. Then we generate random numbers
        $\mathcal R_0$, which are uniformly distributed between 0 and 1, and
        solve $\mathcal R_0=M(<r)/M(<r_{\rm T})$ to get the radial distance.
        Finally, we randomly sample orientations $(\theta,\phi)$, which are
        uniformly distributed on a sphere, and the tuples of $(r, \theta,
        \phi)$ can specify the positions of particles.

    \item {\bf Sampling particle velocity:} The sampling of particle velocities
        relies on Eddington's formula, which is numerically calculated and
        tabulated at first\footnote{We refer to Prof. Martin Weinberg's lecture
            note for the detailed derivation in
        \url{https://courses.umass.edu/astron850-mdw/eddington.pdf}.}. We first
        generate three random numbers $\mathcal R_1$, $\mathcal R_2$ and
        $\mathcal R_3$, which are uniformly distributed between 0 and 1. Then
        we can get the radial velocity $v_{\rm r}$ and tangential velocity
        $v_{\rm t}$ through solving
        \begin{align}
            \mathcal R_1 &= \frac34\left[\frac{v_{\rm r}}{v_{\rm max}} -
            \frac13 \left(\frac{v_{\rm r}}{v_{\rm max}}\right)^3\right] +
            \frac12\\
                \mathcal R_2 &= \frac{v_{\rm t}^2}{v_{\rm max}^2 - v_{\rm r}^2}
            \end{align}
            where $v_{\rm max}(r) = \sqrt{2(E_T - V(r))}$ is the maximum
            velocity. Then we evaluate
            \begin{equation}
                \frac{f\left(v_{\rm r}^2/2 + v_{\rm t}^2/2 +
                V(r)\right)}{f\left(V(r)\right)} >\mathcal R_3,
            \end{equation}
            then we accept the velocity tuple $(v_{\rm r}, v_{\rm t})$ if the
            above assertion is true and otherwise discard it. Finally, we
            generate an orientation on the plane perpendicular to $\mathbf{\hat
            r}$ for the tangential velocity component.

\end{itemize}

\section{Impact of binning on NFW profile fitting}%
\label{sec:impact_of_binning_on_nfw_profile_fitting}

\begin{figure*}
    \centering
    \includegraphics[width=1\linewidth]{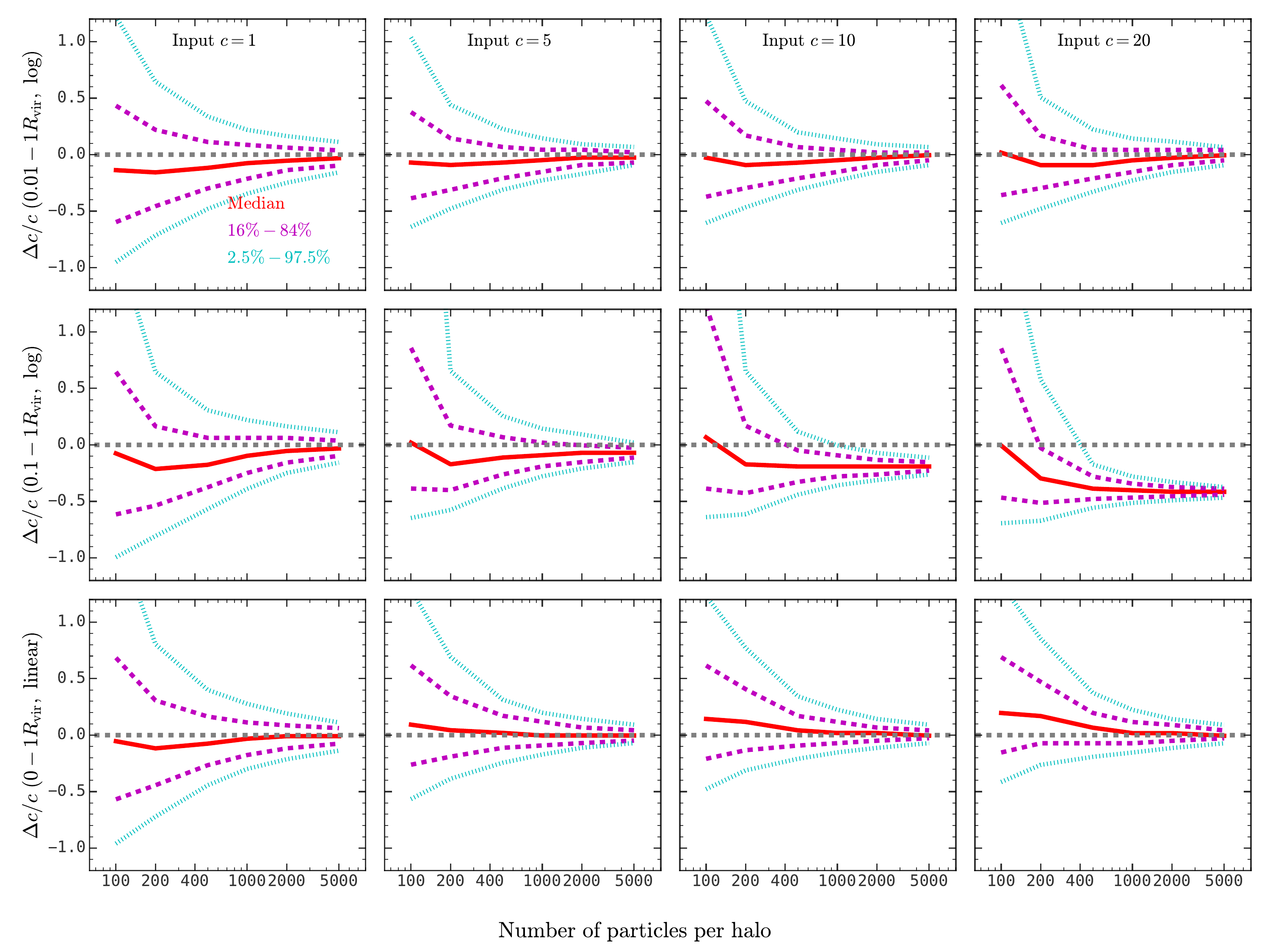}
    \caption{
        Similar to Fig.~\ref{fig:figures/conc_ideal_test}, except that
        different rows are the results obtained from the NFW profile fitting
        method with different binning strategies: binning on a logarithmic
        scale within $0.01-1r_{\rm vir}$ (top panels), binning on a logarithmic
        scale within $0.1-1r_{\rm vir}$ (middle panels), and binning on a
        linear scale within $0-1r_{\rm vir}$ (bottom panels).
    }%
    \label{fig:figures/conc_fitting_binning_test}
\end{figure*}

Fig.~\ref{fig:figures/conc_fitting_binning_test} shows the fractional
concentration deviation for the NFW profile fitting method on ideal NFW halos
generated from \textsc{HaloFactory}, and the three rows correspond to the
following binning strategies
\begin{enumerate}
    \item $(0.01-1r_{\rm vir}, ~\rm log)$: 20 equally spaced radial bins on a
        logarithmic scale from $0.01r_{\rm vir}$ to $r_{\rm vir}$, which is
        used in the main body of the paper);
    \item $(0.1-1r_{\rm vir}, ~\rm log)$: 20 equally spaced radial bins on a
        logarithmic scale from $0.1r_{\rm vir}$ to $r_{\rm vir}$
        \citep[e.g.][]{bhattacharyaDarkMatterHalo2013};
    \item $(0-1r_{\rm vir}, ~\rm linear)$: 20 equally spaced radial bins on a
        linear scale from $0$ to $r_{\rm vir}$
        \citep[e.g.][]{chenRelatingStructureDark2020};
\end{enumerate}
As one can see, the bias in the first and last binning methods converges to
zero as the particle number increases, but the second binning method
underestimates the halo concentration, especially for high-concentration halos.
This happens because the second method does not fit the inner density
distribution. As the particle number decreases to a few hundred, all three
methods produce systematics at the level of $\approx 10-20\%$, and the sign of
the systematics depends both on the binning method and the concentration of the
input halo. These results demonstrate that it is not straightforward to find an
optimal binning to achieve a reliable estimation of the halo concentration,
particularly when individual halos are only sampled by a small number of
particles.

\section{Other methods to estimate halo concentration}%
\label{sec:other_methods_to_estimate_halo_concentration}

\begin{figure*}
    \centering
    \includegraphics[width=1\linewidth]{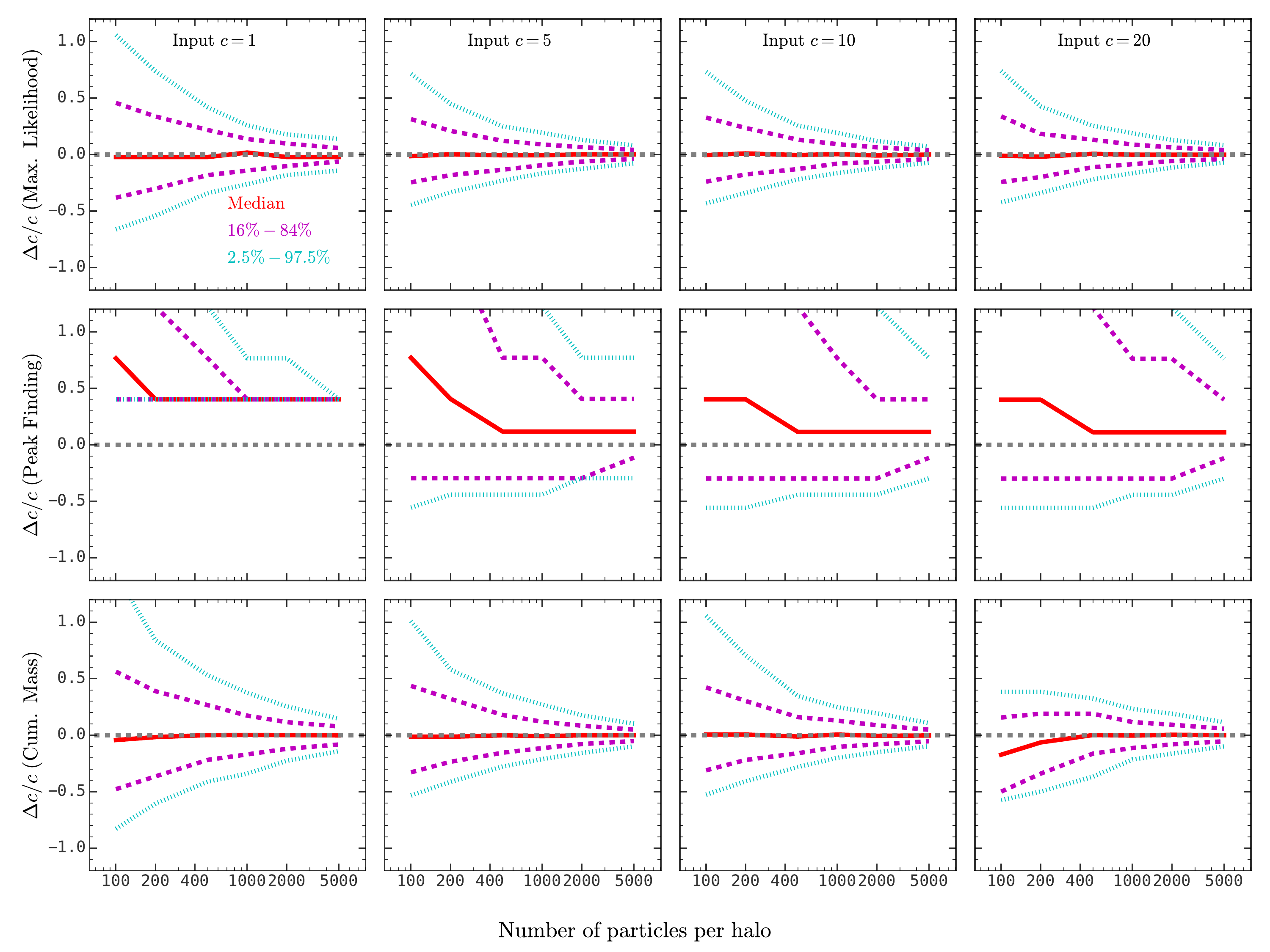}
    \caption{
        Similar to Fig.~\ref{fig:figures/conc_ideal_test}, except for three
        other methods to estimate halo concentration: the maximum likelihood
        method (top panels), the peak finding method (middle panels), and the
        cumulative mass method (bottom panels).
    }%
    \label{fig:figures/other_methods}
\end{figure*}

\paragraph*{Maximum likelihood:} We propose to use the maximum likelihood
method to estimate the concentration parameter of NFW halos. To begin with, the
NFW differential mass profile is given by
\begin{equation}\label{eq:diff_mass_p} {\mathrm d M(<r)\over \mathrm dr} =
    {M_{\rm vir}\over r_{\rm vir}}{x\over m(c)(c^{-1} + x)^2}\,.
\end{equation}
where $m(c) = \ln(1 + c)  - c / (1 + c)$, $x = r/r_{\rm vir}$. Then, for a set
of particles, each represented by the normalized radial location
$x_i=r_i/r_{\rm vir}$, one can express the logarithmic likelihood function as
\begin{equation}
    \ln\mathcal L \propto \sum_i\left[\ln(x_i)-\ln m(c) - 2\ln(c^{-1} +
    x_i)\right]\,.
\end{equation}
The concentration parameter can then be estimated by maximizing this likelihood
function. We note that the maximum likelihood method is unambiguous and
applicable to different halo profiles. However, the summation in the likelihood
function needs to be evaluated in each iteration of the optimization, which
makes this method computationally expensive, especially for finely sampled
halos.

\paragraph*{Peak finding:} The NFW differential mass profile (see
equation~(\ref{eq:diff_mass_p})) peaks at $r = r_{\rm s}$, so the concentration
parameter $c=r_{\rm vir}/r_{\rm s}$ can be inferred by locating the peak of
equation~(\ref{eq:diff_mass_p})
\citep[e.g.][]{childHaloProfilesConcentrationMass2018}. In order to suppress
the noise for poorly sampled halos, we first smooth the differential mass
profile with a three-point Hanning filter, which is
\begin{equation}
    f(r_i) = \frac14\left[f(r_{i - 1}) + 2  f(r_{i}) + f(r_{i + 1})\right]\,.
\end{equation}
Here we calculate the NFW differential mass profile with 20 bins from
$0.01r_{\rm vir}$ to $r_{\rm vir}$ on a logarithmic scale, and the two bins at
both ends are dropped after the smoothing step. In this case, the peak is
constrained within $0.014r_{\rm vir}$ and $0.713r_{\rm vir}$, so that the
minimum and maximum concentrations that can be recovered are $1.4$ and $70.3$.
We note that the result obtained here is subject to the choice of binning.

\paragraph*{Cumulative mass:} For ideal NFW halos, the half-mass radius
$r_{\mathrm h}$, which encloses half of the total halo mass, is given by
\begin{equation}
    \frac12 = {g(r_{\rm h})\over g(r_{\rm vir})}\,,
\end{equation}
where $g(x) = \ln[(r_{\rm s} + x) / r_{\rm s}] - x/(r_{\rm s} + x)$. Therefore, the concentration
parameter can be inferred through $r_{\rm s}$ by numerically solving this equation
\citep[e.g.][]{langVoronoiTessellationNonparametric2015}.


Fig.~\ref{fig:figures/other_methods} shows the performance of these three
methods of concentration estimation on ideal NFW halos. Firstly, the maximum
likelihood method performs the best, even outperforming our $R_1$ method (see
Fig.~\ref{fig:figures/conc_ideal_test}). This method can be easily generalized
to other profiles. However, the optimization of the likelihood is
computationally expensive, especially for finely sampled halos, since the
summation of the logarithmic likelihood of all particles needs to be evaluated
during each optimization iteration. Secondly, the peak finding method makes no
assumption about the specific form of the halo profile. However, this method
performs poorly with large systematics and scatter, since it requires the
evaluation of the (differential) density profile, which poses challenges to the
choice of binning: a large bin size will not sample both ends well, while a
small bin size will cause large shot noise. Finally, the cumulative method
performs quite well, except for the $\approx 20\%$ systematics for
poorly sampled high-concentration halos. Moreover, this method requires
numerically solving a non-linear equation, whose complexity is comparable to
the conventional NFW profile fitting method.

\section{Uncertainties of the $R_1$ method}%
\label{sec:uncertainties}

\begin{figure*}
    \centering
    \includegraphics[width=1\linewidth]{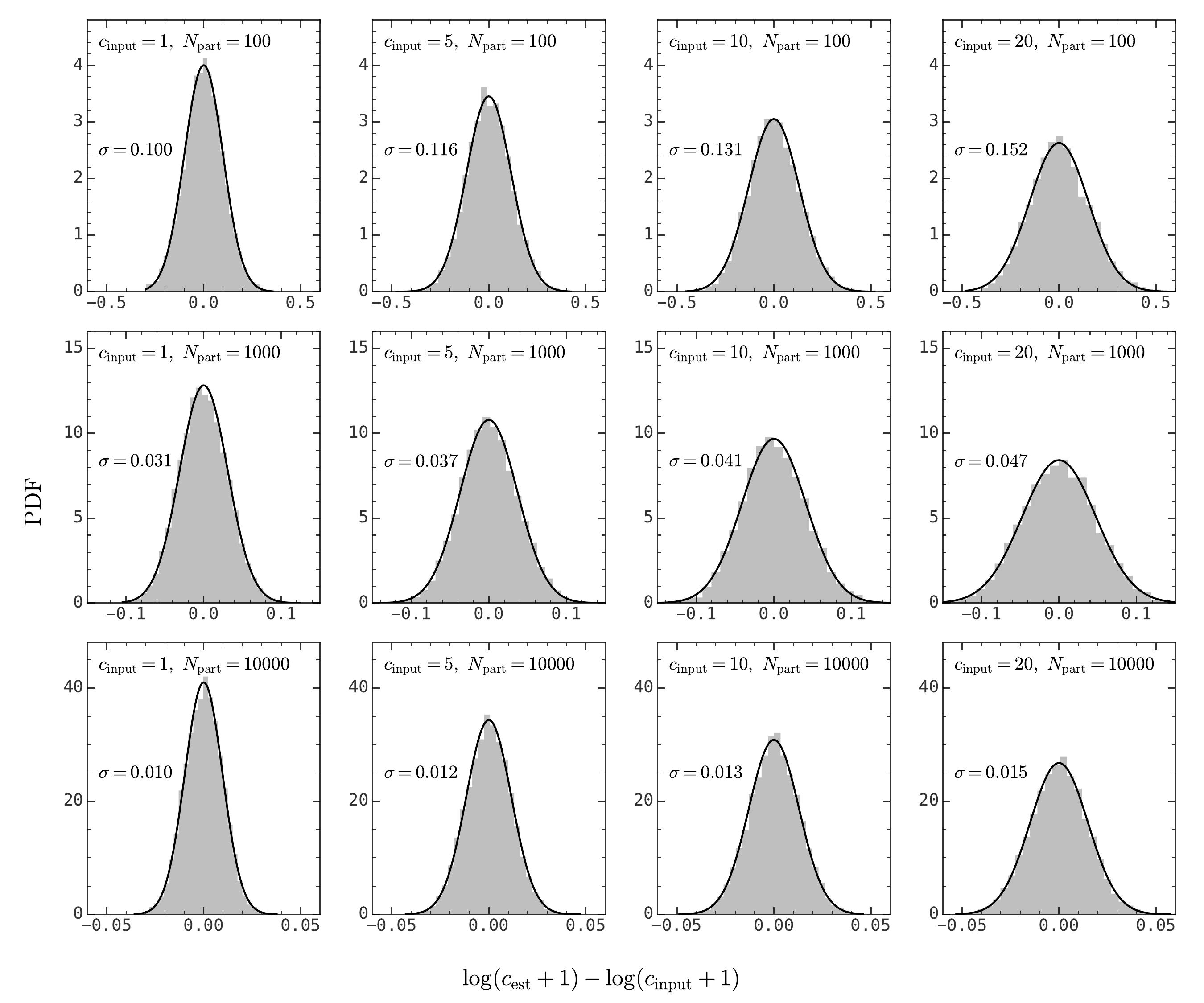}
    \caption{
        Probability distributions of the deviation of the concentration
        estimated using the $R_1$ method, $c_{\rm est}$ from the input
        concentration, $c_{\rm input}$ for different $c_{\rm input}$ (different
        columns) and different numbers of particles (different rows); note the
        different $x$-axis ranges. In each panel, the distribution of $\log
        (c_{\rm est} + 1) - \log (c_{\rm input} + 1)$ is shown as a  gray
        histogram, and the standard deviation, $\sigma$, calculated with
        equation~(\ref{eq:sigma}) is indicated. A Gaussian distribution of
        $\mathcal N(\mu=0, \sigma=\sigma)$ is shown in black solid line. This
        figure shows that the logarithmic deviation of the estimated
        concentration from the true concentration can be described by a
        Gaussian distribution.
    }%
    \label{fig:figures/conc_err_distribution}
\end{figure*}

\begin{figure}
    \centering
    \includegraphics[width=0.9\linewidth]{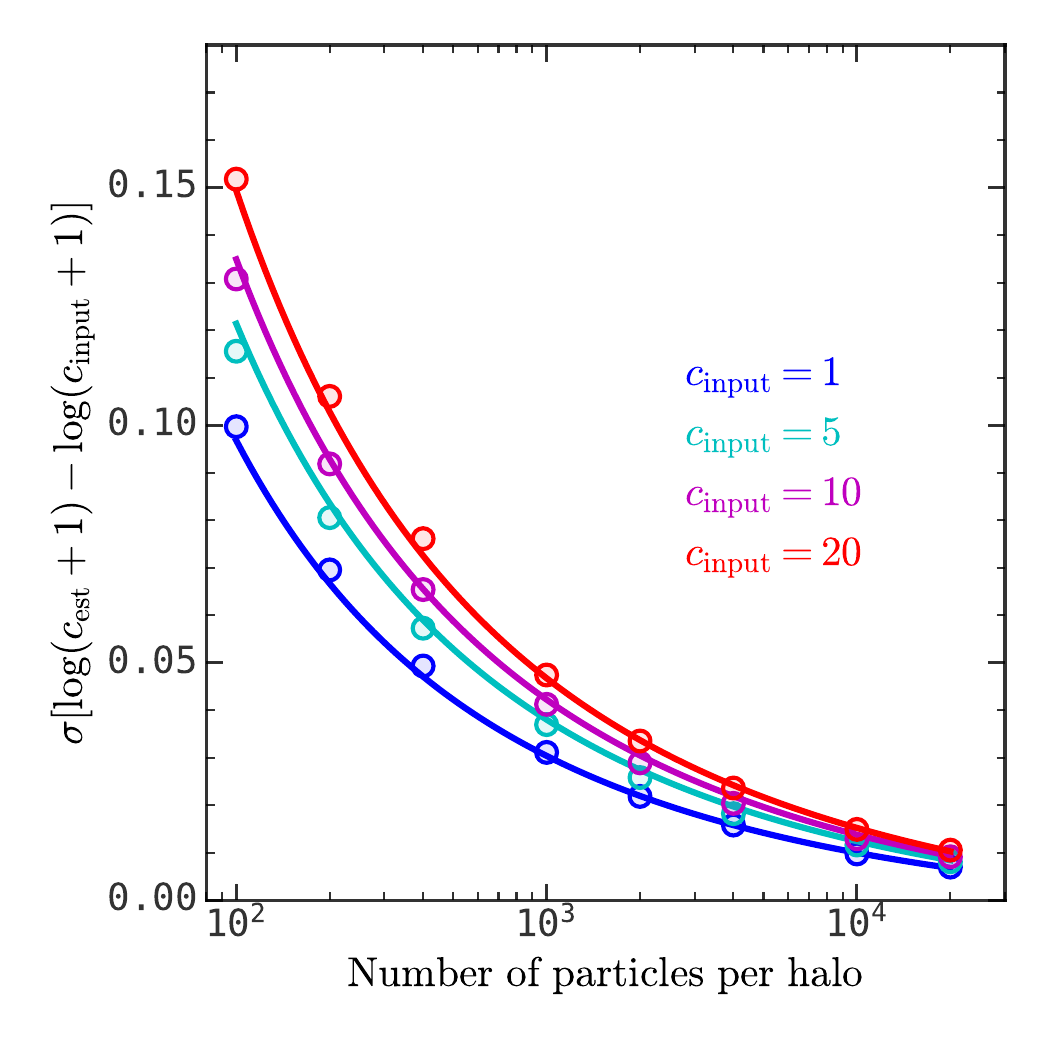}
    \caption{
        The standard deviation of $\log (c_{\rm est} + 1) - \log (c_{\rm input}
        + 1)$ as a function of particle number per halo for different input
        concentrations, where $c_{\rm est}$ is estimated with out $R_1$ method.
        The circles are results obtained with equation~(\ref{eq:sigma}) for ideal
        NFW halos. The solid lines are the fitting results of
        equation~(\ref{eq:sigma_fitting}).
    }%
    \label{fig:figures/sigma_relation}
\end{figure}

Here we quantify the statistical uncertainties of our $R_1$ method in
estimating halo concentration for ideal NFW halos.
Fig.~\ref{fig:figures/conc_err_distribution} shows the distribution of the
logarithmic deviation of the estimated concentration with our $R_1$ method from
the true concentration of ideal NFW halos. For each input concentration and
particle number, we generate 10,000 realizations of NFW halos and estimate their
concentration based on equations~\ref{eq:r1} and \ref{eq:r1_c}. The logarithmic
deviation is shown in the gray histogram. We then calculate the standard
deviation of the logarithmic deviation as follows:
\begin{align} \label{eq:sigma} \sigma &= \sqrt{\frac{1}{N_{\rm
    samp}}\sum_{i=1}^{N_{\rm samp}}\Delta_i^2}\\ \Delta_i &= \log(c_{{\rm est},
i} + 1) - \log(c_{\rm input} + 1)\nonumber
\end{align}
where $N_{\rm samp}=10,000$ in this case. Finally, we over-plot the probability
distribution function of a Gaussian distribution with a mean of 0 and a
standard deviation of $\sigma$ in each panel. Here one can see that this
Gaussian distribution matches the histogram very well.

Fig.~\ref{fig:figures/sigma_relation} shows the relation between $\sigma$ and
the sampling particle number for four different input concentrations. Here one
can see that $\sigma$ is about $0.10-0.15$ dex for halos sampled with only 100
particles. With increasing sampling, the uncertainty of the concentration
estimation rapidly decreases, to a value of 0.01 dex when the number of
particles reaches $2\times 10^4$. We also find that the dependence of $\sigma$
on the particle number and the input concentration can be fitted with
\begin{align} \label{eq:sigma_fitting}
    \sigma(c, N_{\rm part}) &= \frac{a_1(c)}{(\log N_{\rm part})^2} + \frac{a_2(c)}{\log N_{\rm part}}\\
    a_1(c) &= 0.330c + 0.515\nonumber\\
    a_2(c) &= -0.063c -0.095\nonumber
\end{align}
where $N_{\rm part}$ is the particle number. And this function is over-plotted
in Fig.~\ref{fig:figures/sigma_relation} as solid lines.

\section{Density profile for low-mass and high-concentration halos}%
\label{sec:outlier}

\begin{figure}
    \centering
    \includegraphics[width=1\linewidth]{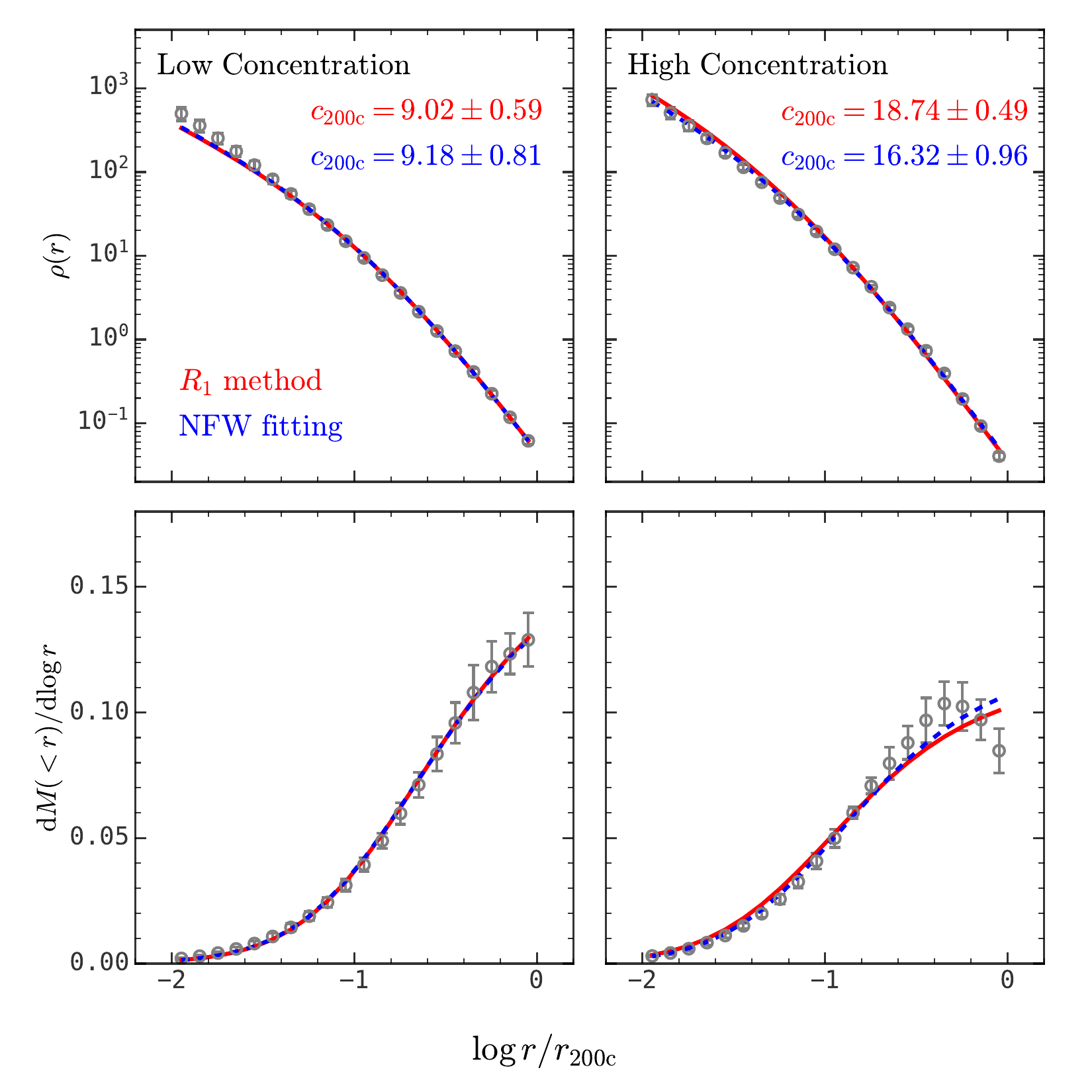}
    \caption{
        The density distribution of low-concentration (left panels) and
        high-concentration (right panels) halos with $11\leq \log(M_{\rm
        200c}/[h^{-1}{\rm M_\odot}]) < 11.5$ selected from TNG50-1-Dark. The
        top panels and bottom panels show the density distributions and the
        mass distributions, respectively. The red solid and blue dashed lines
        show the NFW profile from the mean concentration of each subsample,
        where the concentration is estimated with our $R_1$ method and the NFW
        fitting method, respectively. This figure demonstrates that the
        discrepancy of concentration estimation from the NFW fitting method and
        our $R_1$ method for low-mass and high-concentration halos in
        Fig.~\ref{fig:figures/conc_compare_r1_fitting_tng100_1_3} is attributed
        to the deviation from the NFW distribution due to the stripping of the
        halo outer regions.
    }%
    \label{fig:figures/high_conc_outlier}
\end{figure}

In order to understand the discrepancy in the concentration estimated from the
NFW fitting method and our $R_1$ method for low-mass high-concentration halos
in Fig.~\ref{fig:figures/conc_compare_r1_fitting_tng100_1_3}, we select
high-concentration halos with $11\leq \log(M_{\rm 200c}/[h^{-1}{\rm M_\odot}]) <
11.5$ from TNG50-1-Dark and plot their density distribution in the right panels
of Fig.~\ref{fig:figures/high_conc_outlier}. For comparison, we also plot the
density distribution of low-concentration halos in the same mass range (left
panels). Here one can see that the mass distribution of high-concentration
halos deviates from the NFW profile due to stripping in the outskirts of halos.
The density and mass distributions reconstructed from both NFW fitting and
$R_1$ methods match the data equally well, despite the
$(18.74-16.32)/18.74\approx 13\%$ systematics between their results.

\section{Impact of cosmology on the mass-concentration relation}%
\label{sec:cosmology}

\begin{figure}
    \centering
    \includegraphics[width=0.9\linewidth]{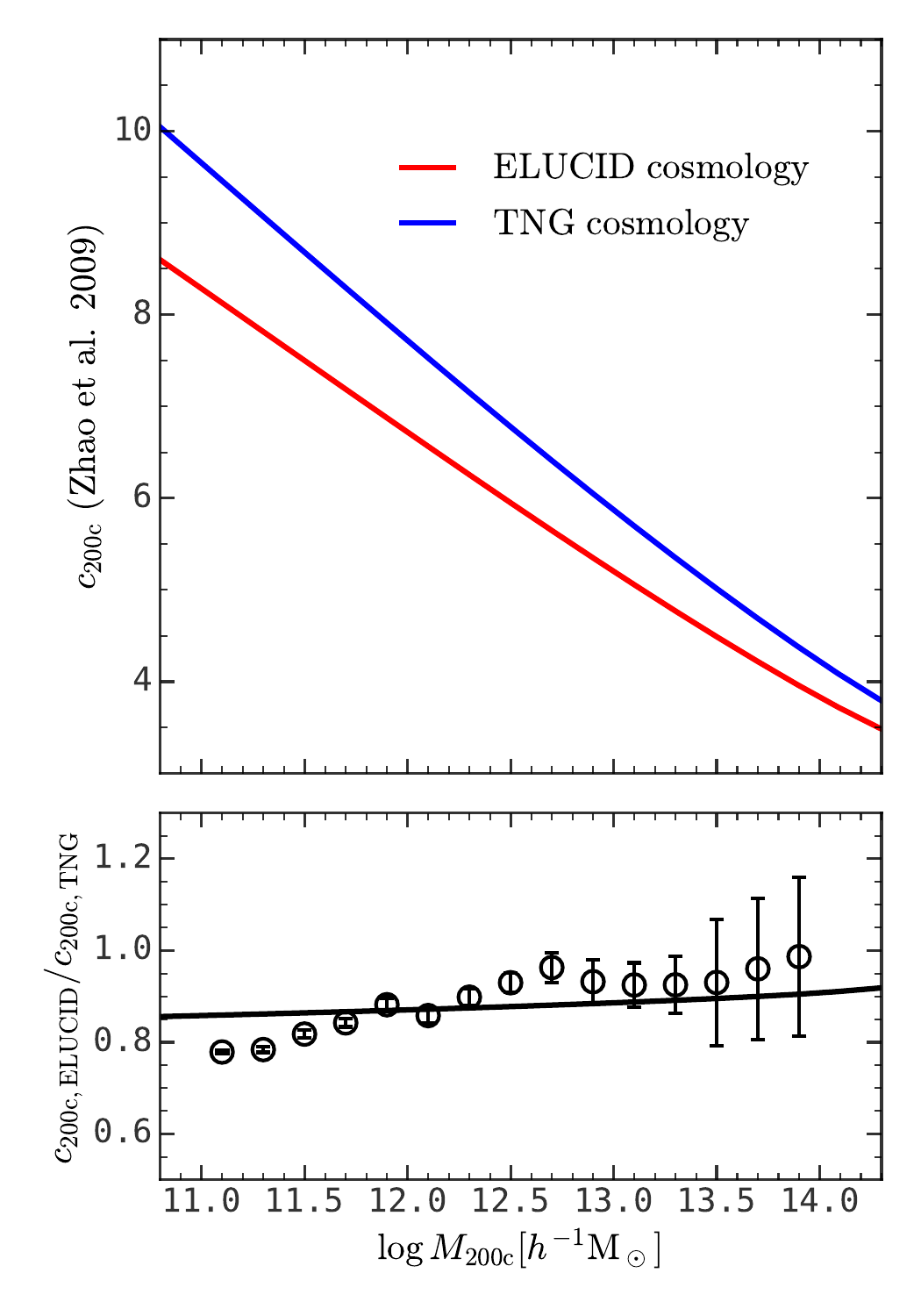}
    \caption{
        Top panel: The mass-concentration relation generated from the
        analytical model in \citet{zhaoACCURATEUNIVERSALMODELS2009} with the
        cosmological parameter values used in TNG100-1-Dark and ELUCID (see
        \S\,\ref{sec:data}). Bottom panel: The ratio between the
        mass-concentration relations from two different sets of cosmological
        parameters. The black solid line is the prediction of the model from
        \citet{zhaoACCURATEUNIVERSALMODELS2009}, and the black error bars are
        the result obtained from simulations. The difference of the
        mass-concentration relation in TNG100-1-Dark and ELUCID can be
        attributed to the difference in their cosmological parameters.
    }%
    \label{fig:figures/conc_cosmology}
\end{figure}

Fig.~\ref{fig:figures/conc_cosmology} shows the mass-concentration relation for
different cosmological parameters generated from the analytical model in
\citet{zhaoACCURATEUNIVERSALMODELS2009}. Here one can see that the cosmology
adopted by TNG yields higher concentrations than that of ELUCID at a given halo
mass, as we see in Figs.~\ref{fig:figures/mass_concentration_relation_tng} and
\ref{fig:figures/mass_concentration_relation}. The bottom panel shows the ratio
between these two analytical relations as the black solid line, and the error
bars show the ratio between the mass-concentration relations for simulated
halos in TNG100-1-Dark and ELUCID. The two ratios are broadly consistent with
each other, indicating that the difference in the mass-concentration relation
between TNG100-1-Dark and ELUCID is caused by the different cosmological
parameters they adopt.

\section{Estimating $c_{\rm 200c}$ from $c_{\rm 500c}$}%
\label{sec:c500c}

\begin{figure}
    \centering
    \includegraphics[width=\linewidth]{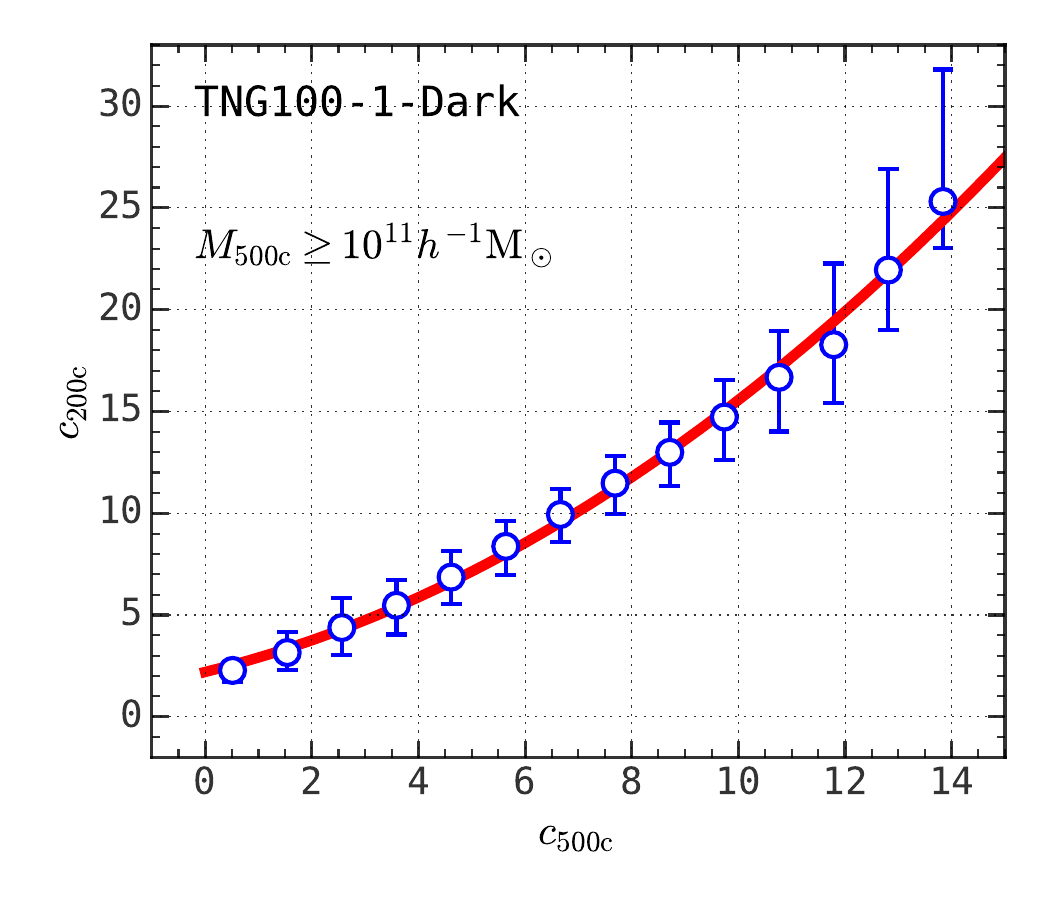}
    \caption{
        The relation between the concentrations obtained from
        equations~\ref{eq:r1} and \ref{eq:r1_c} with $r_{\rm vir} = r_{\rm
        200c}$ and $r_{\rm vir}=r_{\rm 500c}$. The blue error bars show the
        $16^{\rm th}-50^{\rm th}-86^{\rm th}$ percentiles and the red solid
        line is the fitting function in equation~(\ref{eq:c500}).
    }%
    \label{fig:figures/mass_concentration_tng_500c}
\end{figure}

Fig.~\ref{fig:figures/mass_concentration_tng_500c} shows the relation between
the concentrations obtained from equations~\ref{eq:r1} and \ref{eq:r1_c} with
$r_{\rm vir} = r_{\rm 200c}$ and $r_{\rm vir}=r_{\rm 500c}$, which are the
radius within which the mean density is 200 and 500 times the critical density,
respectively. And the corresponding halo masses are $M_{\rm 200c}$ and $M_{\rm
500c}$. The relation between $c_{\rm 200c}$ and $c_{\rm 500c}$ is fitted with
\begin{align}
    \label{eq:c500} c_{\rm 200c} &= g_1c_{\rm 500c}^2 + g_2c_{\rm 500c} +
    g_3\,,\\
    g_1 &= 0.070,~~g_2 = 0.638, ~~g_3 = 2.202\,,\nonumber
\end{align}
which has marginal dependence on halo mass. With this relation, we can infer
the concentration parameter $c_{\rm 200c}$ by integrating equation~(\ref{eq:r1})
only to $r_{\rm 500c}$.

\section{$V_{\rm max}$ estimation for pre-infall halos}%
\label{sec:_v__rm_max_estimation_for_pre_infall_halos}

\begin{figure*}
    \centering
    \includegraphics[width=\linewidth]{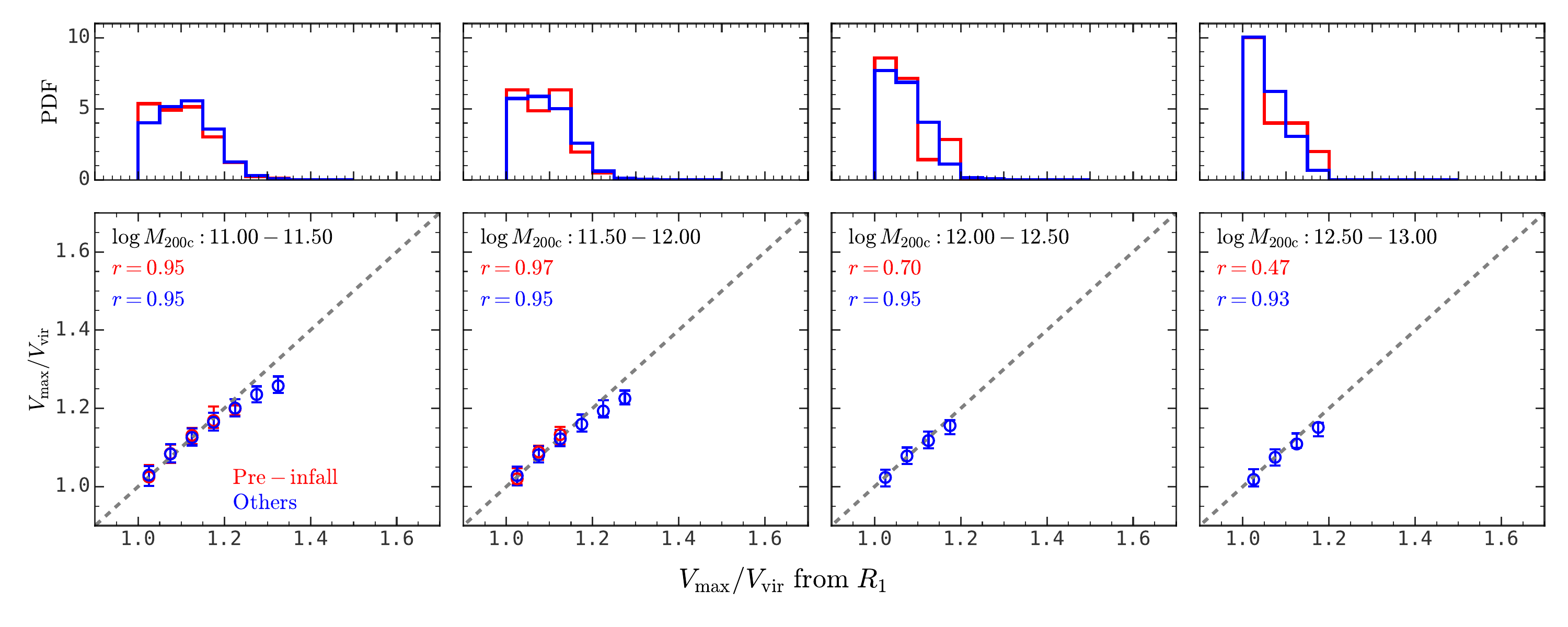}
    \caption{
        Similar to Fig.~\ref{fig:figures/vmax_compare_r1_fitting_tng100_1_3},
        except that these halos are selected from TNG100-1-Dark at $z= 0.7$.
        Red error bars are for halos that will be accreted by other halos and
        become a satellite subhalo at the next snapshot, and blue error bars
        are for the remaining halos. The environmental effects before accretion
        do not affect the estimation of $V_{\rm max}/V_{\rm vir}$ from $R_1$.
    }%
    \label{fig:figures/vmax_compare_r1_preinfall_tng100_1_3}
\end{figure*}

Fig.~\ref{fig:figures/vmax_compare_r1_preinfall_tng100_1_3} compares $V_{\rm
max}/V_{\rm vir}$ calculated from equation~(\ref{eq:vmax}) and estimated from
$R_1$ for halos selected from TNG100-1-Dark at $z= 0.7$. Here halos are
divided into two subsamples: ``pre-infall'' halos are those that will be
accreted onto other halos to become subhalos in the next snapshot; ``others"
are halos excluding ``pre-infall" halos. No additional systematics are seen for
pre-infall halos, indicating that environmental effects on super-halo scales
have insignificant effects on the estimation of $V_{\rm max}/V_{\rm vir}$ from
$R_1$.

\bsp	
\label{lastpage}
\end{document}